%% file: main.tex
\documentclass[fleqn,usenatbib]{mnras}

\usepackage{newtxtext,newtxmath}
\usepackage[T1]{fontenc}

\DeclareRobustCommand{\VAN}[3]{#2}
\let\VANthebibliography\thebibliography
\def\thebibliography{\DeclareRobustCommand{\VAN}[3]{##3}\VANthebibliography}

\usepackage{graphicx}	% Including figure files
\usepackage{adjustbox}
\usepackage{bm}
\usepackage{amsmath}
\usepackage{xcolor}% Advanced maths commands
\usepackage{anyfontsize}

\newcommand\dir{\pmb{\theta}}
\newcommand\nob{\hat{n}}
\newcommand\dob{\hat{\delta}}
\newcommand\dg{\delta_{\text{g}}}

\def\be{\begin{equation}}
\def\ee{\end{equation}}

\title[Clustering systematics corrections]{Joint inference 
 of multiplicative and additive systematics in galaxy density fluctuations and clustering measurements}

\author[F. Berlfein et al.]{
Federico Berlfein,$^{1,2}$\thanks{fberlfei@andrew.cmu.edu}
Rachel Mandelbaum,$^{1,2,4}$
Scott Dodelson,$^{1,2,4}$
Chad Schafer$^{2,3,4}$
\\
$^{1}$ Department of Physics, Carnegie Mellon University, Pittsburgh, PA 15213, USA \\
$^{2}$McWilliams Center for Cosmology, Department of Physics, Carnegie Mellon University,
Pittsburgh, PA 15213, USA\\
$^{3}$Department of Statistics \& Data Science, Carnegie Mellon University, Pittsburgh, PA 15213, USA \\
$^{4}$ NSF AI Planning Institute, Carnegie Mellon University, Pittsburgh, PA 15213, USA
}

\date{Accepted 2024 June 6. Received 2024 June 5; in original form 2024 January 22}

\begin{document}
\label{firstpage}
\pagerange{\pageref{firstpage}--\pageref{lastpage}}
\maketitle

\begin{abstract}

Galaxy clustering measurements are a key probe of the matter density field in the Universe. With the era of precision cosmology upon us, surveys rely on precise measurements of the clustering signal for meaningful cosmological analysis. However, the presence of systematic contaminants 
% I commented out what is below: there is a crucial distinction between systematic contaminants and survey properties, as it's only through some error or problem that the survey properties become systematics - so we should not equate them.
%(also commonly known as survey properties) 
can bias the observed galaxy number density, and thereby bias the galaxy two-point statistics. As the statistical uncertainties get smaller, correcting for these systematic contaminants becomes increasingly important for unbiased cosmological analysis. We present and validate a new method for understanding and mitigating both additive and multiplicative systematics in galaxy clustering measurements (two-point function) by joint inference of contaminants in the galaxy overdensity field (one-point function) using a maximum-likelihood estimator (MLE). 
% Commented out second sentence, first sentence seems to be enough.
%We show how a maximum likelihood estimation (MLE) approach correctly \chad{Maybe a word other than ``correctly,'' which seems a little vague.} \rachel{I would say `accurately' is the statistical terminology for what we mean here}
%estimates systematic contamination in the galaxy overdensity field and can be used to correct for the clustering signal directly at the two-point level.
%\rachel{The previous two sentences seem very repetitive of each other.  Could we perhaps combine them into one sentence, like `We present and validate a new method for understanding and mitigating these systematics in galaxy clustering measurements (two-point function) by identifying and characterizing contaminants in the galaxy overdensity field (one-point function) using a maximum-likelihood estimator.'  I'm afraid otherwise the abstract will be above the MNRAS word limit.} \federico{Combined both and commented out second sentence}  
We test this methodology with KiDS-like mock galaxy catalogs and synthetic systematic template maps. We estimate the cosmological impact of such mitigation by quantifying uncertainties and possible biases in the inferred relationship between the observed and the true galaxy clustering signal. 
% This was a hint to the bias-variance tradeoff which is an important concept that drove our thinking.
Our method robustly corrects the clustering signal to the sub-percent level and reduces numerous additive and multiplicative systematics from $1.5 \sigma$ to less than $0.1\sigma$ for the scenarios we tested. %\chad{Is it clear what
%$\sigma$ means here?} \rachel{This is commonly used to refer to the statistical uncertainties in astro papers, so it's fine.} 
%   Note that fig 4 showed we are doing better than $0.1\sigma$, and this is important: existing methods can already do $1\sigma$-level correction, and most people want to do better than that (so systematics are well below statistical errors).  That's why I modified the text from $1\sigma$ to $0.1\sigma$.
In addition, we provide an empirical approach to identifying the functional form (additive, multiplicative, or other) by which specific systematics contaminate the galaxy number density.  Even though this approach is tested and geared towards systematics contaminating the galaxy number density, the methods can be extended to systematics mitigation for other two-point correlation measurements.
%\rachel{The abstract should probably lead with the science (i.e., we want to enable precise cosmological analysis of clustering) then say what we do.  Also, crucially, the abstract should have some quantitative statement of the result (i.e., not just that we tested it, but we tested it and found we could reduce numerous additive and multiplicative systematics from $X\sigma$ contamination to $Y\sigma$ contamination.} \federico{[CORRECTED]}
\end{abstract}

\begin{keywords}
methods: data analysis – methods: statistical – surveys – cosmology: observations – large-scale structure of Universe.
\end{keywords}

\input{Introduction}
\input{Background}
\input{Methods}

\input{Data_and_Software}
\input{Results}

\input{Discussion}

\input{Conclusion}

\section*{Acknowledgements}

%\rachel{Federico: you will need to collect your co-authors' funding acknowledgments.}\federico{Added Scott, waiting for Chad} 
FB and RM were supported in part by a grant from the Simons Foundation (Simons
Investigator in Astrophysics, Award ID 620789) and in part by the Department of Energy grant DE-SC0010118. SD is supported by U.S. Dept. of Energy contract DE-SC0019248 and by NSF Award Number 2020295. CS is
supported by NSF Award Number 2020295.

Some of the results in this paper have been derived using
the \texttt{healpy} and \texttt{HEALPix} packages. %\rachel{Perhaps the link to SLICS should go under the data availability statement.}\federico{Addressed}
%%%%%%%%%%%%%%%%%%%%%%%%%%%%%%%%%%%%%%%%%%%%%%%%%%
\section*{Data Availability}

%\rachel{Federico: you need to fill in this section.}\federico{Addressed: Added SLICS here, and statement on requesting data}
 
We would like to thank Joachim Harnois-Deraps for making public the SLICS mock data, which can be found at \url{http://slics.roe.ac.uk/}. Any additional data and software generated by the authors is available upon request.

%\rachel{Should review the references carefully and fix any issues.  For example, there appear to be two identical Aihara et al papers listed.  So either the same paper was cited twice with two different BibTeX codes, or two different papers are showing up the same in the PDF due to misformatted BibTeX.  Typically the paper should have a volume and a page number; is the page number missing from the BibTeX?  This issue shows up for a few other refs.  For example, there are two citations to Elvin-Poole, both to papers in PRD in 2018, but the journal is written out differently in the two cases and one of them lacks a page number.  Should check whether they are the same paper cited two different ways (in which case resolve to the one with complete information) or two different papers, and one of them has incomplete BibTeX.  I suggest using ADS for all references because the BibTeX it provides is complete and well-formatted.  Finally, I see at least one ref to an AAS talk rather than a paper (Sanchez et al.) which needs to be fixed to point to a paper. }\federico{Addressed: Aihara was two papers (data release 1 and survey design), Elvin-poole was the same paper with 2 citations, sanchez was replaced by citation recommended by Scott (DES y3 results from 3x2)}



\bibliographystyle{mnras}
\nocite{*}
\bibliography{main} % if your bibtex file is called example.bib

%\begin{thebibliography}{99}
%\end{thebibliography}

%%%%%%%%%%%%%%%%%%%%%%%%%%%%%%%%%%%%%%%%%%%%%%%%%%

%%%%%%%%%%%%%%%%% APPENDICES %%%%%%%%%%%%%%%%%%%%%

\appendix \label{appendix}

\section{Correlated Systematics}\label{appendix: corr_syst }

In order to transform our template maps into an orthogonal space, we construct the pixel covariance matrix of  our template maps:
\be \label{eq: cov_matrix}
C_{ij} = \langle \delta_{t_i}\delta_{t_j} \rangle
\ee
%\rachel{Do we really need the subtraction of the mean?  Overdensities have a mean of zero by definition.}\federico{Yes good point, I removed it.}
We then define the rotation matrix from the eigenvectors of the covariance matrix: {$\mathbfss{C} = \mathbfss{R}\mathbfss{D} \mathbfss{R}^T$}. %\rachel{For notational clarity, matrices should appear different than scalars - see MNRAS author guide.}\federico{Addressed}
 We define the uncorrelated systematic maps as
\be \label{eq: map_rot}
\delta_t^{'} = \mathbfss{R}^T \delta_t
\ee
and the one-point function in equation~\eqref{eq: mult_syst} can be reformulated as 
\be \label{eq: mult_rot}
a_1\delta_{t_1} + ... + a_N\delta_{t_N} = a^T\delta_t = a^T \mathbfss{R} \mathbfss{R}^T \delta_t = a^{'T} \delta_t^{'}
\ee
where $a^{'T} = a^T \mathbfss{R}$. This means that we can fit for the parameters $a_{\text{rot}}$ as if the systematics are uncorrelated and then transform back the parameters by applying the rotation matrix again: $a^T = a^{'T} \mathbfss{R}^T$. We do the same for the $b$ parameters. The corrected estimate of the two-point function is then
\be \label{eq: corr_w_rot2}
w_{\text{corr}} = \frac{\hat{w} - \sum_i^{N_{\text{sys}}} a_{i}^{'2}\langle\delta^{'}_{t_i}\delta^{'}_{t_i}\rangle }{1 + \sum_i^{N_\text{sys}} b_{ i}^{2'}\langle \delta^{'}_{t_i}\delta^{'}_{t_i}\rangle} 
\ee
Alternatively, the observed clustering signal in the original systematics space would be:

%\be
\begin{align*}
\hat{w}(\theta) &= w (\theta)  \left(1 + \sum_i^{N_{\text{sys}}} b_i^2 \langle \delta_{t_i}\delta_{t_i} \rangle  +  \sum_i^{N_{\text{sys}}} \sum_{j \neq i}^{N_{\text{sys}}} b_i b_j \langle \delta_{t_i}\delta_{t_j} \rangle \right)\\ 
& + \sum_i^{N_{\text{sys}}} a_i^2 \langle \delta_{t_i}\delta_{t_i} \rangle  +  \sum_i^{N_{\text{sys}}} \sum_{j \neq i}^{N_{\text{sys}}} a_i a_j \langle \delta_{t_i}\delta_{t_j} \rangle
\end{align*}
%\ee
% Note, you cannot put an align inside an equation - align already is an equation environment.
and the estimated correction becomes:
\be \label{eq: corr_w_crossterms}
w_{\text{corr}} = \frac{\hat{w} - \sum_i^{N_{\text{sys}}} a_i^2 \langle \delta_{t_i}\delta_{t_i} \rangle  -  \sum_i^{N_{\text{sys}}} \sum_{j \neq i}^N a_i a_j \langle \delta_{t_i}\delta_{t_j} \rangle}{1 + \sum_i^{N_{\text{sys}}} b_i^2 \langle \delta_{t_i}\delta_{t_i} \rangle  +  \sum_i^{N_{\text{sys}}} \sum_{j \neq i}^{N_{\text{sys}}} b_i b_j \langle \delta_{t_i}\delta_{t_j} \rangle}
\ee

If working with correlated systematics and choosing not to use an orthogonal set of template maps, mitigation is prescribed by equation~\eqref{eq: corr_w_crossterms}, where the unbiased cross terms are 
\be \label{eq:debias}
a_i a_j = \hat{a}_i \hat{a}_j - \text{Cov}[\hat{a}_i, \hat{a}_j].
\ee

%\section{Generating Systematic Templates}

%Plots on how we fit for the seeing in HSC to approximate power spectra. How the power spectra from all 5 families look like and their respective correlation function on the $100 \; \text{deg}^2$ patch we use.

\section{Parameter Noise and Debiasing}\label{appendix: debiasing }

\begin{figure}
  \center
  
  \includegraphics[width=0.45\textwidth]{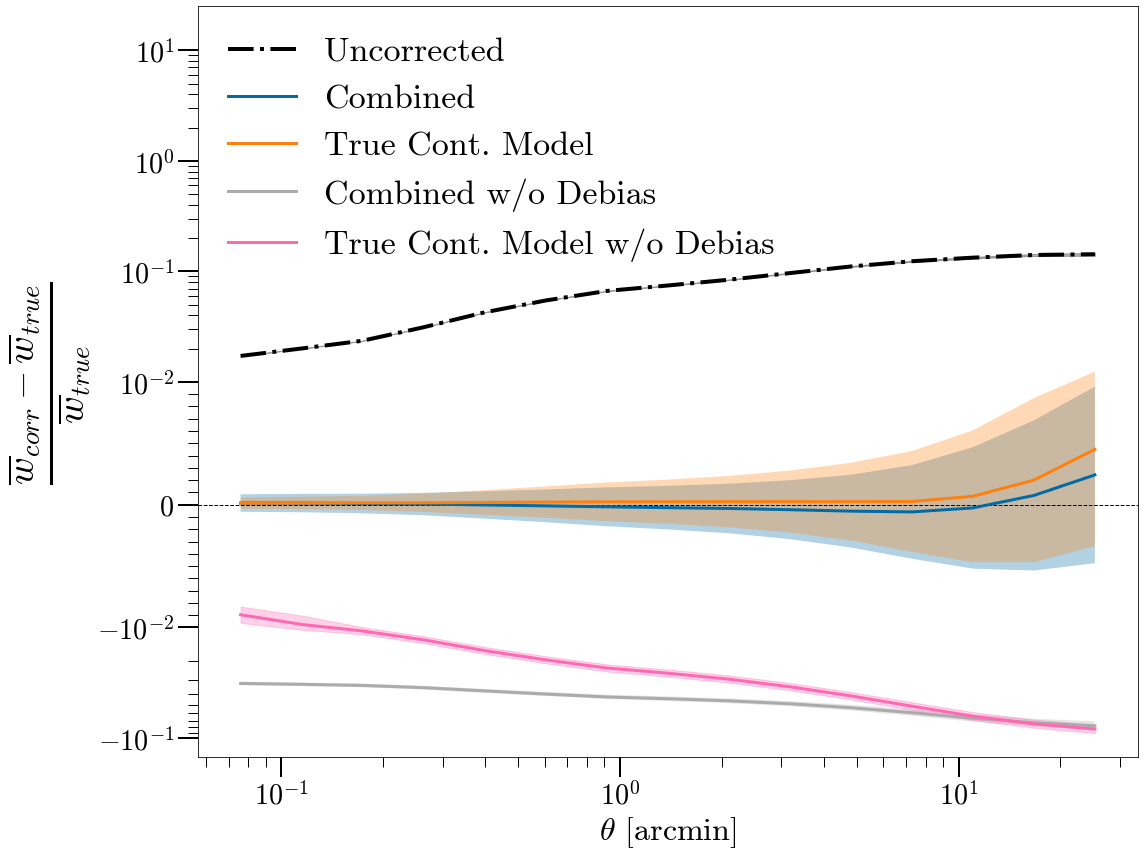}

  \caption{Fractional difference between the average corrected and true two-point function in the presence of 25 systematic contaminants, 10 of which are additive and 15 multiplicative. The plot shows similar information to that of Fig.~\ref{fig: 25syst_correction}, but now comparing the correction method with and without parameter debiasing. The plot shows the importance of the debiasing step, as neglecting it produces a biased (and overcorrected) estimate of the corrected two-point function.  }\label{fig: corr_wo_debiasing }
\end{figure}

We emphasized throughout the paper the importance of debiasing the estimated contamination parameters in order to obtain an unbiased correction of the two-point function. Here we show the resulting correlation function if this debiasing step is ignored altogether. Fig.~\ref{fig: corr_wo_debiasing } shows the results obtained from Sec.~\ref{results: mult_syst} if the debiasing step is omitted. As we can see, omitting this step results in a percent-level bias in the corrected clustering signal -- with an opposite sign, and a magnitude that is approximately 30\% of the original uncorrected systematic. This result further shows the importance of this step in the mitigation method presented in this work.

%%%%%%%%%%%%%%%%%%%%%%%%%%%%%%%%%%%%%%%%%%%%%%%%%%

% Don't change these lines
\bsp	% typesetting comment
\label{lastpage}
\end{document}

%% file: Introduction.tex
\section{Introduction} \label{introduction}

Over the past few decades, groundbreaking observations and theoretical advances have allowed us to construct a remarkably detailed picture of the properties and evolution of the Universe \citep[e.g.,][]{Weinberg_2013}. %\rachel{I'm not familiar with this reference, but it's a conference proceeding that ADS says has only been cited once since it appeared in 2008, so I suspect it might be unrefereed (like many conference proceedings). I suggest replacing this with a citation to one or more standard, refereed review articles, such as this one: \url{https://ui.adsabs.harvard.edu/abs/2013PhR...530...87W/abstract} }\federico{Addressed} 
Early-Universe observations from the cosmic microwave background \citep[CMB; e.g.,][]{Planck_2020} %\rachel{Where is this reference from?  I suggest using ADS in general.  Planck is misspelled multiple time in the BibTeX and the reference as it appears in the paper, and the journal name is mis-formatted in the PDF/compiled version, so this BibTeX needs to be fixed. I'm also not sure it's worth citing a 20-year old WMAP paper, which is primarily of historical interest rather than showing the current state of the field.}\federico{Addressed} 
and distance measurements from Type Ia supernova \citep{Riess_1998, Perlmutter_1999, Filippenko} already provide strong constraints on the currently accepted cosmological model, $\Lambda$CDM. To further test this model and its predictions across the full span of cosmic time and with a range of observables,  photometric galaxy surveys %\rachel{This is a review of photometric redshifts, not galaxy surveys (the latter are much broader).  I suggest either removing this citation, or replacing it with one that matches the broad topic of the sentence. 
 %But I suspect it would be appropriate to just remove it.}\federico{Addressed: removed citation} 
 have emerged as a cornerstone of modern cosmological analyses. Recent large-scale structure (LSS) surveys such as the Dark Energy Survey \citep[DES;][]{DES_2005}, Hyper Suprime-Camera Survey \citep[HSC;][]{Aihara_2017}, and Kilo-Degree Survey \cite[KiDS;][]{de_Jong_2012} have shown strong cosmological constraining power. Upcoming and newly-begun surveys such as the Vera C.\ Rubin Observatory Legacy Survey of Space and Time \citep[LSST;][]{LSST_2019},  %\rachel{This is an unrefereed white paper about DESC, not a paper about LSST.  Should use the refereed LSST overview paper: \url{https://ui.adsabs.harvard.edu/abs/2019ApJ...873..111I/abstract}}\federico{Addressed}, 
 Dark Energy Spectroscopic Instrument \citep[DESI;][]{DESI}, \textit{Euclid} \citep{Euclid}, and the \textit{Nancy Grace Roman Space Telescope} High Latitude Survey \citep{Spergel_2015} will enable even more precise tests of the $\Lambda$CDM model in the coming decade. However, with the increasing survey size enabling greater  statistical precision,
 the challenge of identifying and mitigating systematic contamination to provide robust and accurate constraints will be more important than ever.

Galaxy clustering measurements play an important role in cosmological analyses, as they provide information about the large-scale distribution of matter of the Universe, which carries information about the cosmic expansion history. The relation between the galaxy and matter density field, commonly known as galaxy bias, is a key determinant in clustering \citep{Desjacques_2018} that connects the observed galaxy clustering to the underlying matter field at all physical scales. %\rachel{This statement is important, and needs a reference related to clustering alone.  I suggest adding a sentence also noting the existence of galaxy bias as a key determinant of the clustering (i.e., it's not just the large-scale matter distribution) and referring to a galaxy bias paper such as the Desjacques review}\federico{Addressed}. 
Alongside weak gravitational lensing, galaxy clustering represents one of the three observables used in the current state-of-the-art analysis method for cosmological inference from imaging surveys  \citep[3$\times$2pt analysis; e.g.,][]{krause2021dark, DES_Y3}. Clustering measurements depend on the galaxy (over-)density field and have a very high signal-to-noise ratio in large-area surveys, so potential systematic effects can disproportionately affect the observed galaxy overdensity field and have an outsized impact on cosmological constraints \citep{Elvin-Poole}. Consequently, generating tools for understanding and mitigating the impact of systematics on the galaxy overdensity field is paramount in the effort to produce robust cosmological constraints from large-scale structure measurements.

Systematic effects, in the context of imaging surveys, encompass a wide array of observational and instrumental effects \citep{Sevilla_Noarbe_2021} that can systematically perturb the observed galaxy overdensity field, and thereby bias the measured galaxy two-point statistics. These effects can arise from survey-dependent factors such as atmospheric and observing conditions (e.g., seeing, airmass) and detector effects, or survey-independent (astrophysical) factors such as stellar density or Galactic extinction. In addition, systematics-induced modulation of galaxy redshifts across the sky can also bias our measurements of angular clustering \citep{Baleato_2023}. Accounting for all contaminating systematics can often be difficult, resulting in many possible systematic contaminants to consider \citep[e.g.,][]{DES_clustering}. Moreover, for survey-independent systematics, cross-correlation of different surveys will not help detect and remove their impact, motivating a comprehensive and robust approach to their mitigation.

%\federico{Alternate combination of the 2 paragraphs talking about prev. methods + DES Redmagic (currently commented out but can still see them in latex code):}
%% RM: I like this, thanks!  It avoids the repetitive discussion of mitigation methods and focuses on why do we need to keep working on method development.
Multiple methods have been developed for mitigating galaxy overdensity systematics. We discuss these methods in more detail in Sec.~\ref{background:2}. However, despite significant progress in mitigating galaxy overdensity systematics, challenges remain. The DES Y3 clustering results \citep{Rodriguez_Monroy_2022} are a notable illustrative example. Comparison between results from the color-selected redMaGiC sample \citep{Rozo_2016} and the flux-limited MAGLIM \citep{Porredon_2021} galaxy samples, designed for 3$\times$2-pt analysis, revealed unanticipated inconsistencies. This sparked great interest in the area of systematics treatment even though extensive validation determined the cause of the inconsistencies was most likely systematics-dependant sample selection. One pressing difficulty lies in the empirical determination of the form of the systematics contamination to the overdensity field, particularly when multiple systematics effects combine in complex ways. While simulations offer valuable insights, developing empirical methods for modeling and learning about the functional form of systematics contamination is an important path forward for the field.

In this study, we use a new and more generalized modeling approach for systematics contamination of the galaxy overdensity field.  Our approach is empirical: we provide a generalized functional form for the contamination, and a method for constraining its parameters using the real survey data. We apply this method to jointly infer and mitigate contaminants in the clustering signal and validate its performance using KiDS-like mock galaxy catalogs \citep{Harnois2018} with synthetic systematics incorporated into them. Our goal is to provide a method for empirically characterizing and mitigating more complex systematic contamination. 

The structure of the paper is as follows. In Section~\ref{background} we provide background on methods for mitigating galaxy overdensity systematics and the connection to cosmology. In Section~\ref{methods} we outline our new systematics characterization and mitigation method. Section~\ref{data and software} details the mock extragalactic galaxy catolog used for testing our method, along with approaches used in its analysis.  In Section~\ref{results} we present the results of applying our method under a variety of conditions, building up to a multi-systematic case that emulates the real complexity of survey data. Section~\ref{discussion} discusses the implications of our results and possible applications to real data. Finally Section~\ref{conclusions} summarizes our findings and presents the challenges and future work.

%% file: Background.tex
\section{Background}\label{background}
Here we outline the background for this work, starting with the importance of clustering in  cosmology and how systematics can bias cosmological analysis. We then introduce the galaxy overdensity formalism and contamination model in Subsection~\ref{background:1} and discuss various existing methods to deal with systematic contamination in Subsections~\ref{background:2} and~\ref{sss: summary_of_methods}.

\begin{figure*}
    \centering
    \includegraphics[width=0.8\textwidth]{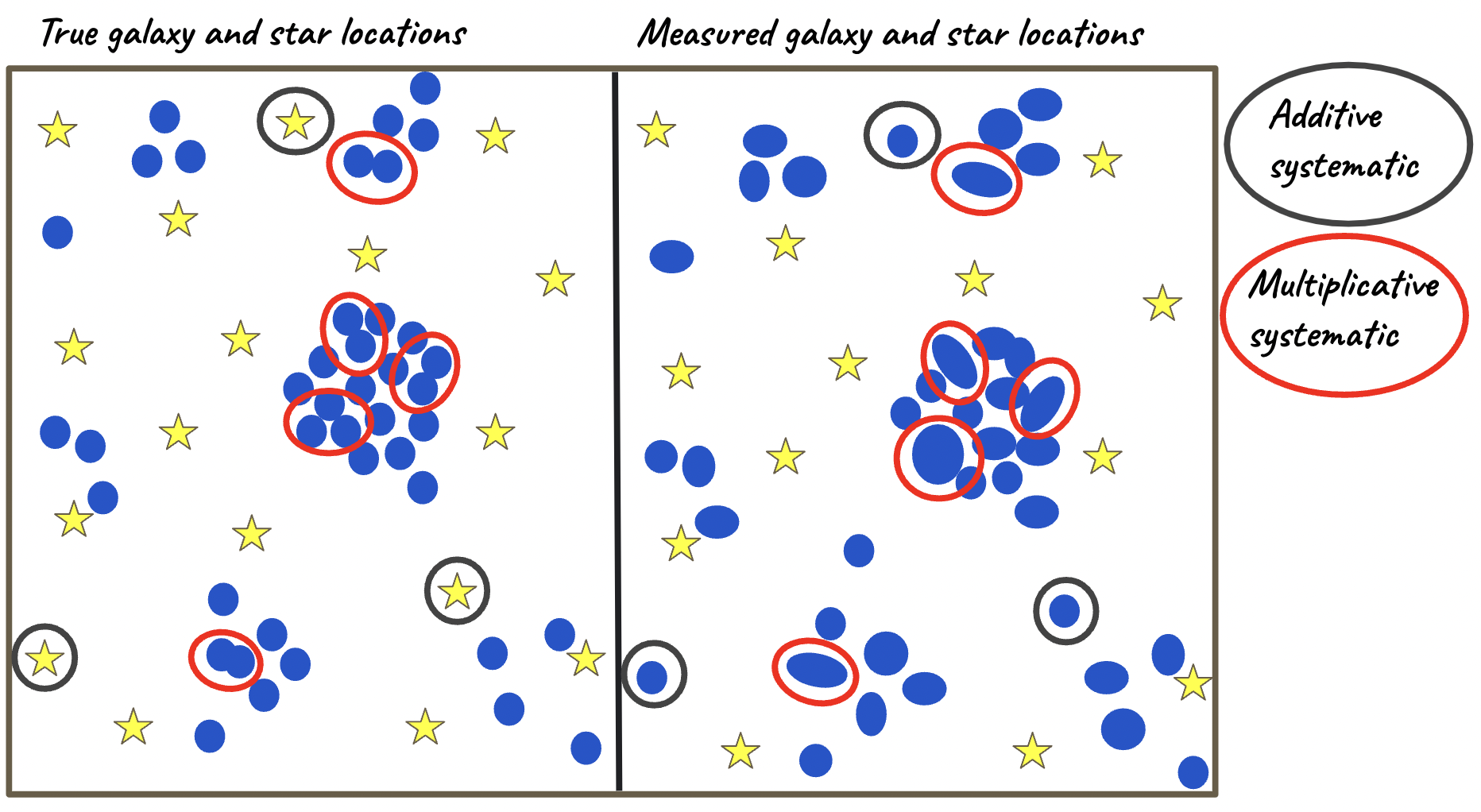}
    \caption{Schematic illustrating examples of additive and multiplicative systematics on the galaxy density field. Yellow stars and blue ellipses represent stars and galaxies, respectively. The left panel shows the true galaxy and star locations, while the right panel represents the measured locations. The regions where the measured and true galaxy distributions differ are circled on both panels, with the color of the circle indicating the type of systematic that is generated. An example of a multiplicative systematic is the unrecognized blending of two or more distinct galaxies in the observed distribution, resulting in a multiplicative effect to the observed number density. An example of an additive systematic is whenstars aremisidentified as galaxies in the measured distribution. This illustration shows how multiplicative systematics depend on the galaxy density field while additive ones are independent of the galaxy distribution. %\rachel{Red and green are not distinguishable for colorblind individuals.  Consider using a different pair of colors from a color blind-friendly palette.}\federico{Addressed, changed to dark grey}
    }
    \label{fig:systematics_drawing}
\end{figure*}

\subsection{Connection to Cosmology} \label{background: cosomology}

The matter distribution of the Universe is of fundamental
interest in cosmology as it carries important information about the evolution and structure of the Universe over time. It is therefore used for testing  theoretical models for cosmic acceleration and understanding the $\Lambda$CDM cosmological model.  %\rachel{Why?}\federico{Adressed} 
Current cosmological analyses, such as the 3$\times$2pt analysis \citep{krause2021dark, DES_Y3}, use the combined information from clustering, shear, and galaxy-galaxy lensing to extract cosmological information (e.g., matter energy density, amplitude of matter fluctuations)  %\rachel{about what?}\federico{Adressed} 
in a Bayesian inference framework. Combining these probes is advantageous because the probes are complementary and combining them enables us to marginalize over observational and theoretical uncertainties associated with single probes. We use galaxies to probe the underlying matter density field, but marginalizing over the galaxy-matter relationship requires modeling both linear and non-linear behavior, with the linear relationship that holds on large scales described as $\dg = b \delta_m \rightarrow w_{\text{gg}} = b^2 w_{mm}$, where $\delta_{\text{g}/m}$ represents the galaxy/matter overdensity field, $w$ the correlation function and $b$ the linear galaxy bias. %\rachel{You have not defined these $w$'s yet.}\federico{Adressed} 

The galaxy bias \citep[for a review, see][]{Desjacques_2018} represents the relation between the galaxy and matter distribution and depends on the mass of the galaxy's host dark matter halo, along with other properties such as luminosity and redshift. Galaxy bias parameters are fitted during the inference process and are crucial in connecting the galaxy clustering signal to the matter clustering. %\rachel{Reference a review article about galaxy bias and give at least one sentence about them depending on the mass of the galaxy's host dark matter halo and potentially other properties. }\federico{Addressed} 
If a simple linear bias model is used, clustering is then sensitive to $b^2$ while galaxy-galaxy lensing is sensitive to $b$ ($w_{\text{g}\kappa} = b w_{m\kappa}$, where $\kappa$ represents the lensing convergence),  %\rachel{Do we need this equation?  If so, need to define $\kappa$}\federico{Adressed}, 
so we can eliminate $b$ by using both probes. Therefore, biased measurements of the galaxy angular correlation function will bias the cosmological inference analysis. It is explicitly shown in \cite{Elvin-Poole} that ignoring systematic contamination to the galaxy clustering signal can cause overestimation of the galaxy bias parameters and underestimation of $\Omega_m$, for example. For this reason, mitigation of systematic contamination in galaxy clustering plays a crucial role in producing an unbiased cosmological analysis. %\chad{This paragraph feels crucial to motivating the
%work. Maybe move it to the top of the section?} \rachel{I very much agree - I nearly wrote the same suggestion at the start of the section.  The only issue is that it uses some notation that would need to be defined in it if we move it up - but that's OK, it's probably worthwhile.  If it's moved, it would probably need a different title, like `Galaxy clustering and cosmology'.}\federico{Adressed} \rachel{One part of this is left to address: you have to make sure you have not made use of notation that is defined later (or if you use it, then define it here).  I see at least two instances of undefined notation, $\delta$ and $b$. In the case of the latter, you refer in words to `galaxy bias' but do not say that $b$ is galaxy bias. Please do a check for this and fix such instances.}\federico{Adressed}

\subsection{Galaxy Overdensity and Contamination Formalism}\label{background:1}

%\rachel{The discussion below is strikingly abstract, without examples of what kinds of systematics might be multiplicative versus additive in the density.  My suggestion is to give an example of each type so people can understand these.  For example, for additive systematics, you could list spurious galaxy detections due to stars mistakenly identified as galaxies (this should depend on the stellar density but not the galaxy density). For multiplicative systematics, you could use any effect that modulates the selection function (detection probability) of galaxies. }\federico{Addressed}

In order to create a reliable contamination model for galaxy overdensity systematics, it is useful to understand what kind of contaminants can be present and how they can affect the galaxy number density. For example, spurious galaxy detections due to stars mistakenly identified as galaxies can have a purely additive effect on the observed galaxy number density: to lowest order, this effect should depend on the stellar density but not on the galaxy density. Spatial variations in the PSF size can modulate the selection function (detection probability) of galaxies, for example due to modifications in the measured signal-to-noise ratio or modulation in the degree of unrecognized blending between galaxies. This effect results in a multiplicative effect on the galaxy number density, as its magnitude depends on the local density of galaxies. An example illustration of additive and multiplicative effects on observed images is depicted in Fig.~\ref{fig:systematics_drawing}. We emphasize that we have only mentioned examples here, and that subtleties in the image processing can give rise to numerous other additive or multiplicative systematics in the galaxy overdensity field.

With these examples in mind, we start by defining notation that describes how contaminants modulate the observed galaxy overdensity field. We model the observed galaxy number density, $\nob_g\!(\dir)$, in a given direction $\dir$ on the sky, as the true galaxy number density $n_g\!(\dir)$ modulated by some systematic contamination function $f_{\text{syst}}(\dir)$ and an additive term $f^{\text{add}}_{\text{syst}}(\dir)$: 
\be \label{eq:ng_background}
    \nob_g\!(\dir) = n_g\!(\dir) (1 + f_{\text{syst}}(\dir)) + f^{\text{add}}_{\text{syst}}(\dir).
\ee
 Since we are interested in the clustering signal quantified as a two-point correlation function of the galaxy overdensity field, we must write equation~\eqref{eq:ng_background}
in terms of the galaxy overdensity field. We write the true number density in the usual way $n_g\!(\dir) = \bar{n}_g (1+ \delta_g\!(\dir))$, where $\delta_g$ represents the overdensity of some galaxy density field and bars over quantities indicate ensemble averages.

If we assume\footnote{This assumption amounts to assuming that the spatially varying additive term $f_\text{syst}^\text{add}$ in equation~\eqref{eq:ng_background} averages to 0 across the survey.  If it did not, then in practice, our estimate of $\bar{n}_g$ would be modified.}  $\bar{n}_g  = \bar{\nob}_g$, equation~\eqref{eq:ng_background} becomes:
\be \label{eq:deltag_background}
\dob_g(\dir) = \delta_g(\dir) (1 +  f_{\text{syst}}(\dir))  +  f_{\text{syst}}(\dir) + \frac{f^{\text{add}}_{\text{syst}}(\dir)}{\bar{n}_g}
\ee
 after dividing by $\bar{n}_g$ and subtracting $1$. Here, too, quantities without vs.\ with hats represent the true vs.\ observed field. We notice two additive terms and one multiplicative bias term. The additive terms $f_{\text{syst}}$ and $f^{\text{add}}_{\text{syst}}/\bar{n}_g$ can be combined into a single term that describes the additive systematics  contamination to the overdensity field. Note that the combined additive terms may differ from the multiplicative term -- they are not, by definition, identical.

% Note: I separated this off into a new paragraph, because the previous part was simply notation, whereas what comes next is our implementation of the method.  It's not always true - e.g., if measuring the galaxy correlation function from a catalog, no pixelization or map-level calculation occurs.

In practice, we can obtain direct estimates of the overdensities and of observable quantities that can produce systematics by pixelizing the observed survey footprint. That is, all observed galaxies can be binned into equal area pixels from which these quantities can be calculated. %\rachel{I'm not sure what is being implied here.  By definition this is true as long as we don't have some additive systematic with a nonzero mean.  Unless you are trying to distinguish between the $\bar{n}_g$ without systematics but with statistical uncertainties, versus the one without statistical uncertainties?  But this whole selection has been about systematics, so this isn't clear to me.}\federico{Addressed} \rachel{I don't agree with the modified text.  It is true that statistical uncertainties can mean the mean density in the survey differs from the true mean in the Universe, but in practice, people use the definition that $\bar{n}_g$ is defined within the survey.  (Deviations from that due to cosmic variance are then treated as part of the covariance matrix.)  There is a good practical reason to do this: the only way we can, in practice, define the mean density is using the data that we actually have.  It seems to me you are trying to define the mean density differently than is commonly done.}\federico{Addressed. I think you are right about the way I'm defining the true number density, so I will delete that part.}

%\rachel{I think  Eq.~\eqref{eq:deltag_background} needs some explanation, like why are there  two additive terms.  Also, a crucial feature of Eq.~\eqref{eq:deltag_background} is that it has a multiplicative bias term, which some methods ignore -- but we haven't used the term `multiplicative' in this section, and I think we should, when describing Eq.~\eqref{eq:deltag_background}.} \federico{Addressed}

\subsection{Methods for Systematics Correction}\label{background:2}

%\rachel{NaMaster is commonly used galaxy clustering code with systematics correction.  Given that it is in quite active use by the community, I think it is worth referencing in what's below (I believe it could fit under section 2.2.1, but you should check). }

Many methods have been developed to address the challenging problem of correcting for systematic contamination in clustering, in response to the need for robust tools to deal with contamination of potentially unknown form. For reference, \cite{Huterer} provide an extensive analysis and a review of multiple decontamination methods, along with their own method using \texttt{ElasticNet}. For consistency we stick closely to the formulations and notation used in \cite{Huterer}. %\chad{Is this the name of their method? If it
%just uses the elastic net approach, I would write it that way.}
Here we provide a summary of some commonly used methods, with a goal of motivating our own.  %\rachel{I think you should have a final part of this subsection that brings everything together - what are the key assumptions the methods typically make (connected back to  Eq.~\eqref{eq:deltag_background}), successes, and challenges of these methods that motivate our own. That can be short, like one paragraph, but I think it's essential for motivating our own method.}\federico{Addressed}

\subsubsection{Mode Projection and Template Subtraction}\label{sss: mode_proj}

%\rachel{The paragraph below starts out by assuming you already know what is mode projection (conceptually and in some detail
%) - for example, it refers to avoiding the difficulty of inverting the whole-map covariance matrix, without saying what is the mode projection method, and what is this difficulty they are trying to avoid.  I would suggest (a) rewriting to include a 1-sentence conceptual description of mode projection at the start and (b) rewriting anything that assumes detailed knowledge of the method.}

Mode projection was introduced by \cite{Press_1992} and is applied to directly correct the galaxy overdensity power spectra ($C_{\ell}$) for the presence of systematic contaminants by assigning infinite variance to the systematic templates, hence modifying the signal covariance matrix. However, this method suffers from the difficulty of needing to invert the modified covariance matrix. \cite{Elsner_2016} proposed an adaptation of this method that avoided this challenge. For a single contaminant template $t$ and some contamination parameter $\alpha$, the contamination model assumes that the additive terms in equation~\eqref{eq:deltag_background} can be written as $f_{\text{syst}}\!(\dir) + f^{\text{add}}_{\text{syst} }\!(\dir)/\bar{n}_g\!(\dir) = \alpha t (\dir)$. The mode projection formalism neglects %\rachel{I am not sure we want to say `We' as this is not our method.  In this section, when defining others' methods, I suggest using phrases like `The mode projection formalism neglects\dots' or other things that make it clear you are describing others' work.  This statement certainly applies to this sentence but possibly others, you should check.}\federico{Addressed} 
the multiplicative term in  equation~\eqref{eq:deltag_background} under the assumption that $ f_{\text{syst}} \delta_g  \ll 1 $, so the overdensity equation becomes  %\rachel{It isn't obvious to me why this is the case.  I think all we can say is that $f_{\text{syst}} + f^{\text{add}}_{\text{syst}}/\bar{n}_g$ is presumed to be given by $\alpha t$?} \federico{Addressed}
\be \label{eq:mode_proj_deltag}
\dob_g(\dir)  = \delta_g(\dir)   +  \alpha t(\dir).
\ee
%\rachel{One issue with your current notation (suppressing direction dependence) is that it's not obvious that $\alpha$ is a constant while $t$ has direction-dependence.  In my mind that makes the equations for those more confusing for those new to this topic, and is an argument for not trying to suppress the direction-dependence.}\federico{Addressed}
The contamination parameter is estimated using the covariance of the template map and the galaxy field and the variance of the template map itself,
\be \label{eq:mode_proj_alpha}
\alpha = \frac{\hat{\sigma}_{\text{tg}}^2}{\sigma_{\text{tt}}^2},% \; \; \;  \sigma_{\text{td}}^2 = \frac{1}{4\pi} \sum_{\ell = 0}^{\infty} (2\ell + 1) \hat{C}_{\ell}^{\text{td}}
\ee
where $\sigma_{\text{tg}}^2$ here represents the covariance of the $t$ and $\dob_g$ maps over the whole footprint. In the full-sky case the covariance is given by
\be \label{eq:mode_proj_cov}
\sigma_{\text{tg}}^2 = \frac{1}{4\pi}\sum_{\ell = 0}^{\infty} (2\ell + 1) \hat{C}_{\ell}^{\text{tg}}.
\ee
%\rachel{I'm pretty confused about the $\sigma$'s here.  The words say they are a covariance, but I'm not sure how to calculate a covariance matrix for entire maps - or do you mean a point estimate, i.e., for tt do you just mean a single number representing the variance of the $t$ map?  Also, I think any statistic involving the t-g correlation/power squared divided by the t-t correlation/power-squared should give $\hat{\alpha}^2$, not $\hat{\alpha}$, just by dimensional analysis.  So overall, I am confused!}\federico{Addressed} \rachel{Thanks, I see you fixed the $\sigma$ definition, but my point about $\alpha$ does not seem to be resolved; Eq.~\eqref{eq:mode_proj_alpha} still seems to me like it should give you $\alpha^2$, not $\alpha$. }\federico{The definition of $\alpha $ I provide here comes straight from lit review in \cite{Huterer}, and I checked in the original paper that it is consistent. However I think it makes sense that it is not squared: $\langle \dob_g t \rangle = \alpha \langle t t \rangle$ from Eq.~\eqref{eq:mode_proj_deltag}, which is consistent with the definition I gave here}
We can write the same expression for $\sigma_{\text{tt}}^2$ by replacing the cross power spectrum with the auto power spectrum of the template maps. Note that in practice the covariances in equation~\eqref{eq:mode_proj_alpha} can be calculated at the pixel-level directly rather than by measuring the power spectra. %\rachel{We cannot measure for the full sky, so we cannot measure all $\ell$ modes.  Do people really try to measure the power spectrum at all $\ell$ and integrate it? Or do they just take the variance of the values in the map directly (which is theoretically related to the integral over the power spectrum for all $\ell$)?}\federico{Addressed} \rachel{This doesn't address my question.  I am asking, in practice, do people really use equation 5 or its tt analog  -- measuring power spectra and integrating them -- or do they simply take the covariance using the measured values in pixels?  The text is unclear on this still.  It focuses on relationships between quantities but not how people actually make the measurements.}\federico{Addressed} 
The method can be straightforwardly extended to the case of multiple systematics, as well as multiple galaxy maps in the case of a tomographic analysis. %\rachel{Do we really have a problem with correlated systematics?  I would have thought the solution for correlated systematics for this method is the same as the one you've used in your paper - is there a fundamental reason that doesn't work?.} \federico{Addressed} 
%\rachel{Should comment on extension to multiple galaxy maps as well, for tomographic analysis.}\federico{Addressed} 
Using this method, a cleaned version of the overdensity field can be obtained from equation~\eqref{eq:mode_proj_deltag}, and the contamination-corrected galaxy power spectrum can be estimated from the resulting cleaned map. %\rachel{I find the previous sentence a little confusing: it seems to conflate correcting the power spectra directly with correcting the map.}\federico{Addressed: Hopefully this correction makes it clear} 
Note that calculating the power spectrum of the cleaned map will result in a biased estimator, and \cite{Elsner_2016} details how to properly debias the estimated power spectrum of the systematics-corrected map.  This method is currently implemented in the open-source galaxy clustering code  \texttt{NaMaster} \citep{Alonso_2019}. It is also worth noting that pseudo-Cl mode projection and ordinary least squares regression at the pixel level have been shown to be equivalent methods \citep{Huterer}.

Template subtraction \citep{Ross_2011,Ho_2012} takes a similar approach to mode projection, keeping the same model contamination of equation~\eqref{eq:mode_proj_deltag}. However, the contamination parameters are calculated for each mode as 
\be \label{eq:tem_sub_alpha}
\alpha_{\ell} = \frac{\hat{C}^\text{tg}_{\ell}}{C^\text{tt}_{\ell}},% \; \; \;  \sigma_{\text{td}}^2 = \frac{1}{4\pi} \sum_{\ell = 0}^{\infty} (2\ell + 1) \hat{C}_{\ell}^{\text{td}}
\ee
and then applied directly to the observed power spectra to obtain a biased estimate of the power spectra:
\be \label{eq:7}
C^\text{gg}_{\ell} = \hat{C}^\text{gg}_{\ell} - \alpha_{\ell}^2 C^\text{tt}_{\ell}.
\ee
\cite{Elsner_2015} details how to properly debias the estimated power spectra in both the full-sky and partial sky limit.

\subsubsection{Iterative Regression and \texttt{ElasticNet} }\label{sss: elastic_net}

A more recent method developed by \cite{Elvin-Poole} uses a simple linear regression approach to iteratively clean the overdensity maps. For a systematic labeled $i$, with template $t_i$, pixels are placed in evenly-spaced bins based on the value of the template within that pixel.  %\chad{Is it clear what ``template value'' means?"} \rachel{No, I think it would be more clear to say something like `For a systematic labeled $i$, with template $t_i$ , pixels are placed in evenly-spaced bins based on the value of the template within that pixel.'}. \federico{Addressed} %\rachel{Not clear what is meant by `the binned quantities'.  So far all we know is that the pixels have been put into bins, but what is the binned quantity that is referred to?  Should this statement come later once the binned quantities have been defined?} 
Using all of the evenly-spaced bins (indexed $j$), regression parameters for each systematic $i$ {$a_i, b_i$} are fit from the equation:
\be \label{eq:8}
\frac{\langle \hat{n}_g   \rangle_j }{\bar{\hat{n}}_{g}} = a_i \langle t_i  \rangle_j + b_i.
\ee
%\chad{The indexing is unclear. What does $i$ refer to?} \rachel{What is $N_\text{obs}$, and how does it differ from previous quantities like $\hat{n}_g$?  Should explain the new notation, or use the old notation if they are actually equivalent.}\federico{Addressed}
Mock catalogs are used to estimate the covariance matrix of the binned quantities needed for fitting.
This method is applied iteratively by identifying the systematics that more highly correlate with the density field and cleaning the map of the more dominant systematics first. It does this by assigning weights to all galaxies based on their pixel template value: $w = 1/(a_i t + b_i)$. It then iteratively proceeds to the next most dominant contaminant performing the same process, which in practice is equivalent to applying a product of weights to the density field.   %\rachel{applying a product of weights for the first two systematics?\federico{Addressed}} 
The process stops once a desired significance level (for the defined null tests) is achieved. %\rachel{Desired significance level in what?}\federico{I was trying to say in relation to the null tests they define, I changed the sentence to say this, hopefully its more clear} 
A final set of weights (the product of each iteration) is then assigned to each galaxy when  calculating the corrected estimate of the two-point correlation function.
%% RM: the next sentences appeared to be copies of ones above, so I commented them out.
%This method is applied iteratively by identifying the systematics that more highly correlate with the density field and cleaning the map of the more dominant systematics first. A final set of weights (the product of each iteration) is then assigned to each galaxy when  calculating the corrected estimate of the two-point correlation function.

Finally, \cite{Huterer} introduced the use of \texttt{ElasticNet}, a linear regression model with L1 and L2 regularization, as a mitigation method. The method uses an additive contamination model to estimate the contamination parameters for each template initially. Then, an additional map-level correction for the templates assumed to be multiplicative is done to account for multiplicative systematics. They also derive the impact of multiplicative systematics on the overdensity covariance and incorporate it into the likelihood. 
 The likelihood for fitting the templates uses a  a contamination model from  equation~\eqref{eq:deltag_background} with $f_{\text{syst}}^{\text{add}} = 0$ and $f_{\text{syst}} \!(\dir) = \pmb{\alpha}\bm{\mathsf{T}} \!(\dir) $, where $\pmb{\alpha}$ consists of a vector of contamination parameters and $\bm{\mathsf{T}}$ is a matrix containing the systematics templates for every pixel. The contamination parameters are fit for by minimizing the loss function
 
 %This method assumes a contamination model from  equation~\eqref{eq:deltag_background} with $f_{\text{syst}}^{\text{add}} = 0$ and $f_{\text{syst}} \!(\dir) = \pmb{\alpha}\bm{\mathsf{T}} \!(\dir) $,  %\rachel{So unlike the others, it works in terms of $n_g$ rather than $\delta_g$?  I didn't realize that.  I find that a little confusing - and below you seem to say it works in terms of $\delta_g$ values, so should we be referring instead  to Eq.~\eqref{eq:deltag_background}?}\federico{Adressed}, 
%where $\pmb{\alpha}$ consists of a vector of contamination parameters and $\bm{\mathsf{T}}$ is a matrix containing the systematics templates for every pixel. 
\be \label{eq: elastic_loss}
\text{Loss} = \frac{1}{2N_{\text{pix}}} ||\dob_g\!(\dir) - \pmb{\alpha}\bm{\mathsf{T}} \!(\dir) ||^2 + \lambda_1 ||\pmb{\alpha}||_1 + \frac{\lambda_2 }{2}||\pmb{\alpha}||_2^2  ,
\ee
where $\dob_g$ represents the observed galaxy overdensity, and $||\cdot||_1$ and  $||\cdot||_2$ are the L1-norm and L2-norm, respectively. These two norms are meant to incentivize sparsity in the estimated parameters (L1-norm) and address possible correlations (L2-norm) between template maps. The correction to the galaxy power spectra then follows a similar procedure to that of mode projection described earlier. However, as previously mentioned, the cleaned maps are estimated by taking into account the multiplicative bias for the templates assumed to have this effect, which was neglected in previous methods. %\rachel{`Additional treatment' seems to imply it does additive and multiplicative -- but the previous text seem to imply it was only multiplicative?}\federico{Adressed} 
The result
is a cleaned version of the overdensity map, which we call here $\hat{\delta}^{\text{clean}}_g$:
\be \label{eq: elastic_cleaning}
\hat{\delta}^{\text{clean}}_g \!(\dir) = \frac{\dob_g \!(\dir) -f_{\text{syst}} \!(\dir)}{1+ f_{\text{syst}}\!(\dir)} .
\ee
%\rachel{The $\hat{s}$ notation is fairly confusing and hard to relate to galaxy overdensities.  Is there any reason not to call it $\hat{\delta}_g^\text{(clean)}$ or something like that?}\federico{Adressed}
When correcting for the impact of systematics on the angular correlation function, a set of weights for each pixel can be produced in a similar manner to iterative regression, where the weight map is given by $\pmb{w}_T = (1+\pmb{\alpha}\bm{\mathsf{T}} )^{-1}$, and $w_{T,j}$ is the weight of pixel $j$. %\rachel{Earlier we said $T$ is a matrix and $\alpha$ is a vector, so it seems like $T\alpha$ is a vector\dots is that right?  The above equation is hard to understand with that in mind.  Should it be written explicitly as a product of factors or something like that?} \federico{Yes $T\alpha$ is a vector, as it represents the weight map, so each element of $w_T$ represents a pixel. This map has the same dimensions as the pixelized overdensity maps. I've added something to hopefully clarify} \rachel{I think what is confusing me is that we're not consistently using the same type of notation for scalars, vectors, matrices (following MNRAS author guide) -- for example we use bold italics for theta, indicating that it's a vector, but alpha is also a vector and it's not bold, so it looks like it is meant to be a scalar.  Similarly $T$ is regular italics so it looks like it should be a scalar, not a matrix.  This is a problem in a few places in the paper but especially in this section I find it confusing.  I suggest reviewing carefully to ensure you are using consistent notations for vectors and matrices.} 
It is important to note that this method does not need mock catalogs, and finds the regularization parameters through a form of cross-validation and partition of the maps.

%\rachel{Federico, please check/confirm the accuracy of the next sentence, which I added in order to help motivate your method:}\federico{Yes looks good, thanks for adding this!} 
While \texttt{ElasticNet} represents an advance in correcting for both additive and multiplicative systematics in the galaxy overdensity, it relies on a priori information regarding which templates are multiplicative without a means for empirically choosing between the two. This becomes a special case of equation~\eqref{eq:deltag_background}, which shows that if there are additive and multiplicative systematics in the number density, the result is {\em different} additive and multiplicative systematics in the overdensity. Therefore, incorrect a priori choices of which templates are multiplicative can lead to biases in the corrected overdensity map.

\subsubsection{Non-linear Methods}\label{sss: NN_methods}

There are a variety of non-linear machine learning methods \citep[e.g.,]{Rezaie, Johnston_2021} that attempt to correct for systematic contamination in galaxy clustering. For example, \cite{Rezaie} uses a similar contamination model as  equation~\eqref{eq:deltag_background},  %\rachel{should this be equation 3, assuming they use overdensity rather than density?}\federico{Adressed}, 
but instead of assuming a specific linear form for the contamination, use a fully connected forward neural network to solve the regression problem. The learned contamination function %\rachel{`selection function' means something kind of different: it's typically a selection probability from 0 to 1, determining how likely it is that you'll select something.  I don't think that's what we mean here, is it?  If not, should use different terminology.}\federico{Adressed} 
can then be applied to weight each pixel and correct the power spectra for systematics contamination. \cite{Johnston_2021} %\rachel{something wrong with the bibtex for is - the ref is showing up strangely (with his first name) in the text}\federico{Addressed} 
uses self-organizing maps \citep{Kohonen} to mimic the systematic-contaminated maps and subtract the spurious modes from the clustering signal.

\subsection{Summary of Methods}\label{sss: summary_of_methods}

%\rachel{This looks like part of section 2.3.3 but I think it is meant to be distinct from nonlinear methods.  I suggest making it a subsubsection called `Summary of methods' or something like that.}\federico{Addressed}
The correction techniques discussed in the previous subsection have achieved notable success in reducing the impact of systematic contamination on clustering measurements and have been applied to a wide range of survey data and simulations. Most of these methods are built on several key assumptions, such as the ability
% removed belowed: to model the systematic effects accurately using a simple additive model, or one where additive and multiplicative terms are the same as each other
to treat systematics contamination as fully additive, or fully multiplicative, rather than a combination of both. While methods with these assumptions have often been sufficient for existing datasets, it is by no means clear that they will be sufficient for datasets with smaller statistical uncertainties.  For example, neglecting multiplicative terms (that is, $\delta_g$ terms in equation~\eqref{eq:deltag_background}) in the correction, or assuming they are the same as the additive term, could result in uncorrected biases \citep{Shafer_2015, Huterer} in cases where those terms exist and differ from the additive systematic term. 

In addition, understanding the existence and level of both the additive and multiplicative systematics terms in the contamination model can potentially help us understand empirically how systematics contaminate the observed galaxy number density. This knowledge could be used to identify areas for improvements in image processing algorithms.

These challenges highlight the need for further advances in correction techniques and modeling that can address the more general model for systematics in galaxy overdensities shown in equation~\eqref{eq:deltag_background}. This motivates us to  develop a generalized and flexible approach that can accurately model and remove systematic contamination in clustering while providing some understanding of the functional dependence on $\delta_g$.
%functional form of the contamination itself. 
This provides a level of interpretability to the results that non-linear ML methods, such as neural networks, often lack.  

%\rachel{I notice that our motivation here is aimed primarily at the methods in 2.3.1 and 2.3.2.  Can we say something about why we don't want to just use a ML method? 
% e.g., interpretability?  If this comment is not easy to address, feel free to ignore it.}\federico{Added a short sentence, let me know what you think.}

%\rachel{Section trails off mid-sentence.}\federico{Yes sorry I started writing the final section last night and was not feeling very inspired. Corrected now. }

%% file: Methods.tex
\section{Methods} \label{methods}

This section describes the method introduced in this work.  First, we introduce the specific contamination model used to quantify and correct for systematic contamination in the one and two-point functions. The likelihood function chosen to characterize the galaxy overdensity field is defined with the inclusion of possible contamination, along with statistical methods to compare different modeling choices. Next we highlight the impact of noise and how to remove biases it introduces in the correction of two-point correlation functions.  Finally, we describe possible selection effects when dealing with real data and mocks.

%\rachel{Section has a bunch of subsections, so you should open with a brief statement of what's happening in this section before diving into the subsections.}\federico{Addressed}

%\rachel{Personally, I think it might be cleaner to distinguish between {\em general background} that defines the notation you'll use, versus the sections where you outline the methodology that is new in this paper.  I would suggest making a `Background' section for the basic parts of section 2.1 defining the notation of galaxy clustering and systematics, for typical methods of correcting the clustering for systematics, and how the clustering connects to cosmology.  Then the methods section can just focus on the way you've chosen to describe systematics and how that connects to correcting the clustering signal. }

%\rachel{This section will need a lot more referencing within it.}

%\rachel{This section has a lot of hard-coded references.  Need to turn them into proper latex references so you can move things around without any problems with the references.}

%\rachel{In the paper and figures, you've called the clustering signal $w$.  But conventionally it's called $w$.  It is important to use the conventional notation (unless you are specifically defining a new quantity) so I suggest changing this throughout text and figures to $w$.} \federico{{RESOLVED}}

\subsection{Systematics Contamination Model} \label{methods: syst_model}

%\rachel{I do not really understand why we are going back to equation 1 here.  Why do we not start with equation 3, which already has the form we want for our combined model?  (i.e., allowing the two terms in the overdensity to differ because of the potential for systematics that are additive vs.\ multiplicative in the density)  To me, this would be the most straightforward use of the background section to define the model.  The current formulation seems unnecessarily convoluted.  I'd like to understand this more before editing the text around here, so I am jumping down to below Eq.~\eqref{eq: gal_overdensity}.}\federico{Adressed}
 We will start from equation~\eqref{eq:deltag_background} and initially consider the case of one systematic, described by some template $t$. In our formulation of the systematics model we choose to define the systematic functions in terms of template overdensities $\delta_t  \!(\dir)$. Similar to other works, we choose $f_{\text{syst}} \!(\dir) = \alpha \delta_t  \!(\dir) $, but now consider $f^{\text{add}}_{\text{syst}} \!(\dir) = \beta \delta_t  \!(\dir)$. So  equation~\eqref{eq:deltag_background} becomes as follows:
 \be \label{eq: gal_overdensity}
\dob_g \!(\dir)  = \delta_g \!(\dir)  + a \delta_t \!(\dir)  + b \delta_t \!(\dir)  \delta_g \!(\dir)
\ee
%\chad{I know that I am being dense here, but I think it would benefit to have a
%more clear statement as to what $\nob_{g,i}$ is defined to be. The way in which 
%the averaging is being performed is unclear to me.}
where $a$ and $b$ are the contamination parameters, absorbing the previous constants: $a = \alpha + {\beta}/{\bar n_g}$ and $ b = \alpha$. This allows us to construct a nested model, for which the cases of purely additive ($b= 0$) and purely multiplicative ($a=b$) number density contamination are special cases.  
%% RM: I added "number density" in the above to avoid confusion.  In fact, there are both additive and multiplicative terms in overdensity in this case; but they are equal because the systematic is purely multiplicative in *number density* (which is not what we're modeling).
Since a given systematic template  might have both additive and multiplicative effects on the number density,  generalization beyond these simple models is  physically motivated %\chad{What does ``such generalization'' refer
%to?}
and allows us to test our assumptions about what contamination models we use. Also note that we have chosen to consider only terms linear in $\delta_t$ since we can ``feature engineer'' through nonlinear transforms of one or more base templates such that the dependence is linear \citep{Elvin-Poole, Huterer}.

%since we do not expect contamination from higher order terms to be significant. 

%In addition, assuming the true galaxy density and size does not correlate with the systematic, when averaging over many regions with the same template value, the observed galaxy density will equal the underlying density modulo the seeing effect. If we look at some arbitrary region, a fixed fraction of the galaxies in that region will pass the resolution cut and be included in the sample.  This basically assumes the galaxy size distribution is similar in over/under-densities.

This formalism can be generalized for the case of $N_{\text{sys}}$ systematics maps, for which we would have:
\be \label{eq: mult_syst}
\dob_g \!(\dir) \approx \delta_g \!(\dir) \left(1 + \sum_i^{N_{\text{sys}}} b_i \delta_{t_i} \!(\dir) \right)  +  \sum_i^{N_{\text{sys}}} a_i \delta_{t_i} \!(\dir).
\ee
From now on we will refer to the case where $\{a,b\}$ are free as the ``Combined'' model, the form when $a = b$ as the ``Multiplicative'' model, and the form when $b = 0$ as the ``Additive'' model, i.e., % Note that you can't use regular quote marks in latex - they show up backwards in the compiled doc.
\begin{equation} \label{eq: syst_models}
\hat{\delta}_g \!(\dir) =
\begin{cases}
    \delta_g \!(\dir) + \sum_i^{N_{\text{sys}}} a_i \delta_{t_i} \!(\dir), \; \;  \text{ Additive} \\     
    \delta_g \!(\dir)(1 + \sum_i^{N_{\text{sys}}} a_i\delta_{t_i} \!(\dir)) + \sum_i^{N_{\text{sys}}} a_i\delta_{t_i}\!(\dir), \; \; \text{Multiplicative}\\
   \delta_g \!(\dir)(1 + \sum_i^{N_{\text{sys}}} b_i\delta_{t_i} \!(\dir)) + \sum_i^{N_{\text{sys}}} a_i\delta_{t_i}\!(\dir), \; \; \text{Combined}
\end{cases}
\end{equation}
As explained in Sec.~\ref{background:2}, most current methods assume either additive or multiplicative contamination models in the galaxy number density. 
% I made this a new paragraph as it seemed unrelated to the previous sentence, so they did not really flow as a paragraph.
If the template maps are uncorrelated with each other and with the underlying true galaxy overdensity distribution, the observed galaxy two-point function would then be: 
\be \label{eq: obs_w}
\hat{w} (\theta) = \langle \dob_g\dob_g\rangle = \langle \delta_g\delta_g\rangle \left[1 + \sum_i^{N_{\text{sys}}} b_i^2\langle \delta_{t_i}\delta_{t_i}\rangle\right] + \sum_i^{N_{\text{sys}}} a_i^2\langle\delta_{t_i}\delta_{t_i}\rangle,
\ee
and our estimate for the corrected two-point correlation function would be 
\be \label{eq: corr_w_uncorrelated}
w_{\text{corr}} (\theta) = \frac{\hat{w} (\theta) - \sum_i^{N_{\text{sys}}} a_i^2\langle\delta_{t_i}\delta_{t_i}\rangle }{1 + \sum_i^{N_\text{sys}} b_i^2\langle \delta_{t_i}\delta_{t_i}\rangle} .
\ee
So in order to correct the clustering signal for systematics, we must estimate the contamination parameters $\{a_i, b_i \}$ and measure the spatial correlation functions of the zero-mean template overdensities. Note that model misspecification can lead to biased corrections of the correlation function, as we will see in Sec.~\ref{results}. Some of our main goals are %\chad{sounds a little
%off since there are three goals listed}%\federico{Addressed}
to introduce a formalism with the flexibility to handle different types of systematics, to robustly correct the two-point function despite the increase in the number of parameters, and to gain an understanding of the functional contamination model empirically from data.

%\rachel{I encourage you to think about where this would go, but I don't believe you've really explained about these maps and how they are determined.  For example, it is not obvious from this text that the $t_i$ are also defined to be mean-zero maps, is it?}\federico{Adressed}

\subsection{Correlated Templates}\label{methods: corr_templates}

The above formalism assumes systematics templates are uncorrelated, but generalizes easily to correlated systematics.  To handle the case of correlated systematics templates, we use the eigenvectors of the template maps' pixel covariance matrix to create an orthogonal set of template maps, $\delta_t^{'}$. Equations~\eqref{eq: obs_w} and~\eqref{eq: corr_w_uncorrelated} can then be rewritten in the rotated frame to mitigate the clustering signal:

\be \label{eq: corr_w_rot}
w_{\text{corr}}(\theta) = \frac{\hat{w}(\theta) - \sum_i^{N_{\text{sys}}} a_{i}^{'2}\langle\delta^{'}_{t_i}\delta^{'}_{t_i}\rangle }{1 + \sum_i^{N_\text{sys}} b_{i}^{'2}\langle \delta^{'}_{t_i}\delta^{'}_{t_i}\rangle} ,
\ee
where $\{ a_{i}^{'}, b_{i}^{'}  \} $ are the contamination parameters in this new space.  %\rachel{Do we really need the `rot' subscripts?  Can we simply use primes, just like for the template maps?  It seems like that would make the notation more transparent.}\federico{Adressed} 
The contamination parameters in the original template space can be retrieved, if needed, by applying the inverse rotation matrix on these new parameters. We want the original parameters back as a way of evaluating parameter estimation and possibly understanding the functional form of contamination from systematics. Details of the derivation of the correlated case are in Appendix~\ref{appendix: corr_syst  }. %\chad{It feels as if this paragraph should
%be headed with {\bf Comment.} or similar.} \rachel{I don't think that is MNRAS style, but I do agree it feels slightly distinct and like something we want to highlight.  Would it be overkill to have this be a short subsection called `Contamination model accounting for correlated systematics templates'?}\federico{It seems a bit too short to have a subsection dedicated to it. What about either: 1) Briefly mentioning that if you wish to retrieve the original parameters refer to the Appendix on details, 2: A more compact statement as a footnote } \rachel{Chad and I were saying everything from `The above formalism\dots' (i.e., the entire paragraph with embedded equation) should be a new subsection.  Not just the three sentences about getting the original parameters back.}\federico{Ok I see, then I think a subsection is ok. I made the change and named it just "Correlated Templates" to keep it short.}

\subsection{Likelihood Function}\label{methods: likelihood_func}

%\rachel{There are some pretty confusing things in this section - previously you used $i$ to index the systematics from 1 to $N_\text{sys}$ but now you are using it to index the pixels.  I think this can cause confusion for readers new to the formalism, and I encourage you to use (throughout the whole paper) one index for systematics and a different index for pixels.}\federico{Adressed}

 To estimate the contamination parameters, we work directly with galaxy and systematic overdensity maps. We use \texttt{healpy} \citep{Healpy} and \texttt{HEALPix}\footnote{http://healpix.sourceforge.net} \citep{Healpix} to pixelize the data and create such maps from galaxy catalogs. Hence, the notation $\delta_{t,j}$ refers to the average value of $\delta_t$ from all galaxies in pixel $j$.  %\rachel{Why is the subscript changing from $t$ to $s$ once we pixelize?  Is this residual old notation?}.\federico{Addressed (yes old notation)}  
 %\rachel{For the sake of clarity and reproducibility, should comment on how we choose the pixel size for this study.}\federico{Addressed} \rachel{It's a good step that you've given the practical considerations you had to weigh.  But for the sake of reproducibility, at some point you need to say the actual pixel size that was used, as well.}\federico{Addressed}. 
 The size of the pixels depends on the sample: it should be chosen such that pixels have a non-negligible number of galaxies, but are small enough to retain information about clustering on the smallest scales of interest. In this work we used pixels of NSIDE = $512$, which have an area of $\sim 47$ arcmin$^2$. %\rachel{In that case, I'm not sure we can justify plotting clustering signals down to 0.1~arcmin, which in any case is smaller than is typically used for science.}\federico{We can adjust the plots to include a vertical line or shaded region showing the scales smaller than the pixel size. A possible justification can be as an exploratory tool: Can we still correct the two-point function at smaller scales than the pixel size? Regarding what is typically used for science, I think we can make some exceptions given the small size of the simulations (10x10 degree box). There are many larger scales I had to cut out due to the size of the box being too small that would be of interest in typically surveys. It seems to me that cutting all scales smaller than the pixel size gets rid of a big portion of the plot. Let me know what you think of all this.} \rachel{We could certainly keep these scales, but then we need to explain the logic: comment on the connection to the pixel size, and to scales used for science, and say why.  A natural place to do so would be in the section where we talk about the clustering signal calculations including the scales chosen.  Then in the results, when commenting on the fidelity of corrections, we can refer back to this issue of correcting below the pixel scale.}\federico{Addressed: I added some comments in Section 4.3 talking about the connection between the pixel scale and the clustering scales. For the comment on the fidelity of the corrections, I will wait to comment something until we discuss the scales we want to use to quantify the bias.} 
 %\rachel{As per our discussion this past week, I think the plan now is to (a) explain why the large pixel scale does not cause an issue for testing the method (because you used them to input the systematics) and (b) mention this question of pixel scale vs.\ characteristic scale of systematics in the future work for applying the method.}\federico{Addressed: point a) is addressed in section 4.3, along with a comment on results section 5.2 by the end of the second paragraph addressing the fidelity of the corrections using this choice. Point b) is addressed in the first paragraph of section 6.2} 
 We use a maximum likelihood estimation approach to estimate the parameters of interest. 
 
 The simplest approach is to model the probability density as a Gaussian distribution, $p(\delta_g) \sim \mathcal{N}(0, \sigma_g)$. The conditional probability distribution $p({\dob_g}|\bm{\mathsf{\delta_t}}, \pmb{a},\pmb{b} )$ for uncorrelated pixels is then:
 
 \begin{align}\label{eq: gauss_likelihood}
& {p_{\text{Gauss}}}({\dob_g}|\bm{\mathsf{\delta_t}}, \pmb{a},\pmb{b}, \sigma_g) \propto \frac{1}{\prod_{j = 1}^N (|1+\pmb{b} \cdot \pmb{\delta}_{t,j}|)}\\
& \quad \times\exp \left\{\frac{-1}{2\sigma_g^2}\sum_{j=1}^N \left( \frac{\hat{\delta}_{g, j} -\pmb{a} \cdot {\pmb{\delta}_{t,j}} }{1 + \pmb{b} \cdot {\pmb{\delta}_{t,j}} }  \right)^2 \right\}, \notag 
\end{align}
where $\bm{\mathsf{\delta_t}} = \{ \pmb{\delta}_{t_1}, ..., \pmb{\delta}_{t_{N_{\text{syst}}}}  \}$ and  $ \pmb{a^T} = \{a_1, ... , a_{N_{\text{syst}}}\}$, $ \pmb{b^T} = \{b_1, ... , b_{N_{\text{syst}}} \}$, $\pmb{\delta}_{t,j}$ is the $j$th column of the matrix $\bm{\mathsf{\delta_t}}$, and $\sigma_g$ is the galaxy overdensity pixel-map variance. The summation goes over all $N$ pixels in the map.  The rows of the matrix $\bm{\mathsf{\delta_t}}$ represent each systematic, while the columns represent the systematic values at a particular pixel. The denominator term is a normalizing factor needed to account for the multiplicative form of rearranging equation~\eqref{eq: mult_syst}.
 In order to parametrize the $\delta_g$ distribution as a likelihood we need to understand what we expect its general form to be. By definition, $\delta_g \geq -1$ and $\langle \delta_g \rangle = 0$, so the distribution is bounded from below but is allowed to take arbitrarily large positive values. This means our distribution will naturally be skewed. For this reason, we also explore modeling the conditional probability of each pixel as a skewed Gaussian distribution. The resulting log-likelihood is: 
%\be 
%\rachel{Note: you have made this an `align*' environment, but also tried to give it a label - that won't work, as any environment with a star in the name does not have a label.  So if you want to refer to this equation later, it should just be an `align' environment.  (Previously you had tried to put the `align*' inside of an equation environment, but that causes a compilation failure - those two cannot be nested.  So I commented out the equation stuff.)  To me this seems like a key equation you want to refer to later, so I suggest making it `align' (to be able to refer to it later in the text) and then using the notag option on one of the two lines.}\federico{Adressed}
\begin{align}\label{eq: likelihood}
& -\ln{\mathcal{L}}(\dob_g|\pmb{\delta_t}, \pmb{a},\pmb{b}, \sigma_g, \gamma) =  \frac{N}{2} \ln\left( 2\pi \sigma_g^2\right) + \sum_{j = 1}^N \ln (|1+\pmb{b} \cdot \pmb{\delta}_{t,j}|) \\
&\quad +\frac{1}{2\sigma_g^2}\sum_{j=1}^N \left(\delta_{g,j}- \xi\right)^2 - \sum_{j=1}^N \ln \left(1 + \text{erf}\left(\frac{\gamma(\delta_{g,j} - \xi) }{\sigma_g \sqrt{2}}\right)\right),\notag
\end{align}
%\ee
where $\delta_g$ contains the in terms $\{{\dob_g},\bm{\mathsf{\delta_t}}, \pmb{a}, \pmb{b} \}$ by rearranging equation~\eqref{eq: mult_syst}, $\gamma$ is the skewness parameter, and $\xi = - \sigma \frac{\gamma}{\sqrt{1 + \gamma^2}}\sqrt{\frac{2}{\pi}}$ is the center of the distribution, in this case defined by assuming that the mean value of $\hat{\delta}_g$ is zero. %\rachel{I'm a little confused by this equation, because $\delta_g$ is defined such that it has a mean of 0.}\federico{Adressed} \rachel{You have deleted the equation I mentioned for the mean overdensity, but have you addressed the root question I was asking?  If our model allows the mean to be non-zero, how does that interact with a real scenario where we always define it to have a mean of zero?}\federico{To clarify, in our particular model we do assume it to have mean zero. The above equation for $\xi$ forces the mean of the distribution to be zero. In the case of the skewed Gaussian, this $xi$ represents the "center" of the distribution}.  %element-wise multiplication. \rachel{If you do element-wise multiplication, you get another vector.  It's not clear to me how you can then add that to $1$ (a scalar).  I think that part of the equation might need some work.} \chad{I suspect
%it should be an inner product instead, but agree that it's confusing.}\federico{Addressed, hopefully this makes more sense now. I have changed the equation to an inner product, and added some extra explanation in the text.} 
The second term in the likelihood serves as a normalization factor. Note that this formula assumes pixels in our galaxy overdensity map are uncorrelated, even though we know this  is not true. However, we show later in Section~\ref{results} that this approximation still enables us to accurately %\chad{maybe a weaker word than ``correctlty?''} \rachel{I think he is probably going for `accurately' (in the statistical sense).}\federico{Addressed} 
mitigate the systematics in the two-point correlation functions. For the case of the combined systematic model with $N_\text{sys}$ templates, we fit $2(N_\text{sys} + 1)$ parameters via the maximum likelihood. Both $\pmb{a}$ and  $\pmb{b}$ consist of $N_\text{sys}$ parameters used in mitigation, while $\sigma$ and $\gamma$ are nuisance parameters describing the overdensity distribution. %\rachel{Don't we still have to fit for them?  i.e., shouldn't we be saying we fit for $2(N_\text{sys}+1)+1$ or $+2$ parameters, with the last 1 or 2 being nuisance parameters?}\federico{Yes we fit for them, that's why its $2(N_\text{sys} + 1)$. The a's and b's are $2N_\text{sys}$ and the nuisance parameters are 2. For example if $N_\text{sys} =2$ , there are 2 a's, 2 b's, 2 nuisance, 6 total or $2(N_\text{sys} + 1)$.}

%\rachel{This is optional, but I think at this point we could start a new subsection called `Distinguishing between likelihood models' -- this content builds on but is distinct from the text that simply defines the likelihoods.}\federico{Yes good point, added.}

\subsection{Distinguishing between likelihood models}\label{methods: distinguishing_likelihood}

To summarize, we have created a nested likelihood model described by equation~\eqref{eq: likelihood}  %\rachel{here's a good place to refer to your likelihood equation}\federico{Addressed} 
that can reduce to a simple Gaussian in the case of $\gamma = 0$, and that reduces to additive or multiplicative systematic terms for particular relationships between $a_i$ and $b_i$ for systematic $i$. This formulation of our model allows us to easily compare models in terms of how well they describe the data using likelihood ratio tests. For shorthand we represent all model parameters by $\Theta = \{\pmb{a},\pmb{b}, \sigma, \gamma \}$.  We define this ratio as
\begin{align}\label{eq: likelihood_ratio}
    \lambda_\text{LR} &= 2\ln \left ( \frac{\max_{ \Theta }\mathcal{L}(\dob_g|\bm{\mathsf{\delta_t}},\Theta)}{\max_{\Theta_0 }\mathcal{L}(\dob_g|\bm{\mathsf{\delta_t}},\Theta_0)} \right )  \\
    &= 2 [\ln\mathcal{L}(\dob_g|\bm{\mathsf{\delta_t}},\hat{\Theta}) - \ln\mathcal{L}(\dob_g|\bm{\mathsf{\delta_t}},\hat{\Theta}_0)], \notag
\end{align}
where $\mathcal{L}$ is the likelihood defined in  equation~\eqref{eq: likelihood},
%\rachel{This equation uses a symbol $\mathcal{L}$ that has never been defined.  Should it be used somewhere in connection with the unnumbered equation above that gives the likelihood (but with a different symbol)?  Note that the arguments in that equation are also different.  Also, while `sup' is standard statistical terminology, it's not standard in astronomy, and needs to be defined.}\federico{Addressed} \rachel{It does not help to say in words that the likelihood is defined in equation 17, because equation 17 defines a completely different symbol ($p$) and gives different arguments for the likelihood as well.  The equations themselves are inconsistent with each other, and this needs to be fixed. }
%\chad{Not to get too technical, but the second equality sort of
%invalidates the need for using $\sup$ in place of $\max$, i.e.,
%if I believe that there is a particular set of parameter values
%at which the maximum is achieved, then the distinction between
%$\sup$ and $\max$ disappears.} \federico{Addressed: I introduce the symbol $\mathcal{L}$ in equation 17 now, which hopefully clarifies any confusion. And replaced sup with max.} 
$\Theta$ and $\Theta_0$ are the unrestricted and restricted parameter spaces, and the hatted variables represent their best fit parameter estimates. So for example, when we want to test whether the additive systematic model is sufficient to describe systematics due to template $t_i$, we can define $\Theta_0 = \{ \Theta : b_i = 0 \}$. Similarly when we want to test the sufficiency of a Gaussian likelihood to describe the data, we can define $\Theta_0 = \{ \Theta : \gamma = 0 \}$. In a likelihood ratio test formulated this way, the null hypothesis is that the data can be adequately described by a simpler model, which usually translates to a model with fewer parameters or restrictions. %\rachel{A model is not a hypothesis, though.  I think the hypothesis is, rather, that the data can be adequately described by the simpler model?}\federico{Addressed} 
If the null hypothesis is true, the quantity $\lambda_\text{LR}$ defined earlier is asymptotically $\chi^2$ distributed. %\rachel{you have not said what is the null hypothesis - should articulate that first, i.e., `In a likelihood ratio test formulated this way, the null hypothesis is \dots.  If the null hypothesis is true, $\theta_\text{LR}$ is \dots}\federico{Addressed}. 
In order to reject the null hypothesis (thereby preferring a more complex model to describe the data), we need to define some critical value $z$ such that we can reject the null hypothesis if $\lambda_\text{LR} > z$. This critical value is often taken to be $z \approx F^{-1}(1-\alpha, r)$, where $F$ is the cumulative distribution function (CDF) of a $\chi^2$ distribution, $\alpha$ is the probability of incorrectly rejecting the null hypothesis,  %\rachel{too vague description}\federico{Addressed}, 
and $r$ the number of degrees of freedom. The value of $\alpha$ is selected based on the desired risk of incorrectly rejecting a simpler model for a more complicated one.    

This test provides an approach to model selection, and hence a general understanding of the validity
of the 
assumptions underlying our models, that takes as default the simpler (null) model, and
hence requires strong evidence to reject that model in favor
of the more complicated alternative. This reflects our general
preference for avoiding overparameterized models. These tests
are enabled by the nested structure of the models under
consideration. However, ultimately we must draw conclusions about the success of our method based on how well they enable us to correct the two-point correlation functions.
%\chad{How about instead
%``This test provides an approach to
%model selection, and hence a greater understanding of the validity
%of the 
%assumptions underlying our models, that takes as default the simpler (null) model, and
%hence requires strong evidence to reject that model in favor
%of the more complicated alternative. This reflects our general
%preference for avoiding overparameterized models. These tests
%are enabled by the nested structure of the models under
%consideration. However, ultimately we must draw conclusions about the success of our method based on how well they enable us to correct the two-point correlation functions.''}
%\rachel{Yes, I think this would better explain the relationship between what we are doing here and later in the paper.}\federico{Addressed: thanks for the suggestion Chad.}

\begin {table} 
\begin{adjustbox}{width=0.7\columnwidth,center}
\begin{tabular}{ ||c|c||} 

\multicolumn{2}{|c|}{Mocks Information} \\ 
\hline
$z$ range & $0.2 < z < 0.4$ \\ [0.1cm]

$N_{\text{realizations}}$ &  119 \\ [0.1cm]

$\overline{N}_{\text{gal}}$ & 303371  \\ [0.1cm]

$\overline{n}_{\text{gal}}$ (arcmin${^{-2}}$) & 0.84 \\ [0.1cm]

Number of Random Galaxies &  3036000 \\ [0.1cm]
\hline
\end{tabular}
\end{adjustbox}
\caption {KiDS-HOD mocks redshift binning used and galaxy number density information. Since we have 119 realizations and one random catalog, number and density information on the mock catalogs is given as an average over all realizations. %\rachel{Do you use the various symbols such as $N_{\text{randgal}}$ elsewhere?  If not, it would be better to simply use words rather than using symbols that are not defined or used elsewhere.  It's fine if the table has to be wider than 1 column to accommodate this.}\federico{Addressed}
}
\label{tab:mocks_description}
\end {table}

\subsection{Noise Debiasing}\label{methods: noise_debiasing}

%\rachel{As we discussed last week, I suggest reframing this completely. Noise bias is a known and well-understood phenomenon, and correcting for it should be articulated as part of the method, rather than as something one does later after finding a bias.  We can then have a single plot that shows the importance of this correction.} \federico{[CORRECTED]}

The correction for biases in the clustering signal in equation~\eqref{eq: corr_w_rot} uses the squares of the estimated contamination parameters ($a^2, b^2$). Under the assumption that our systematics model correctly describes the systematics contamination, the estimated parameter $\hat{a}_i$ can be modeled as the true parameter $a_i$ plus zero-mean noise:   $\hat{a}_i = a_i + n_i$. In this case,  $ \hat{a}_i$ is an unbiased estimate for $a_i$, but its square  $\hat{a}_i^2$ is a biased estimator for $a_i^2$:
\be \label{eq:parm_noise}
\langle \hat{a}_i^2   \rangle = \langle (a_i + n_i)^2   \rangle =  \langle a_i^2  \rangle  + \langle n_i^2   \rangle .
\ee
The unbiased squared terms that should be used in the estimated correction in  equation~\eqref{eq: corr_w_rot} are therefore
\be \label{eq:param_debias}
a_i^2 = \hat{a}_i^2 - \text{Var}[\hat{a}_i],
\ee
where $\text{Var}[\hat{a}_i]$ here represents the noise  variance of the parameter $a_i$.  %\rachel{This explanation is quite confusing in this context, because this is the first time your paper has mentioned realizations, and it's not clear what it's referring to. Since you elaborate below about getting the various, it might be most straightforward here to simply end the sentence at `variance of the parameter $a_i$'.}\federico{Addressed} 
The same considerations and debiasing procedure applies to the $b$ parameters. In the limit of purely additive systematics, the noise debiasing presented here is in essence equivalent to the debiasing methods developed in mode projection and template subtraction.

Note that failure to debias the squared parameters can lead to significant overcorrection of the two-point function. This is shown quantitatively in Fig.~\ref{fig: corr_wo_debiasing } %\rachel{something wrong with this figure reference, it's pointing to the whole appendix rather than to a figure}\federico{Addressed} 
of Appendix~\ref{appendix: debiasing }. This illustrates the necessity of removing the noise bias in the estimate of the squared parameters in order to obtain an unbiased estimate of the clustering signal. This particular debiasing method differs from that of methods like \texttt{ElasticNet}, which deal with the bias-variance tradeoff by using data-calibrated L1 and L2 regularization, as mentioned in Sec. \ref{sss: elastic_net}.  Meanwhile, this approach balances this tradeoff using a direct de-biasing technique similar to that of mode projection and template subtraction.

This approach is straightforward to implement if mock catalogs are available to estimate the variance in the systematics parameters $\{\hat{a}_i, \hat{b}_i\}$. %\chad{should have hats?}\federico{Addressed}. 
However, in the absence of mock catalogs, we will discuss in Section~\ref{discussion} a bootstrapping approach to estimating the noise without the need for mock catalogs. It is important to note that the parameter noise is influenced by multiple factors: 1) the specific form of the likelihood for and power spectra of  $\delta_g$ and $\delta_t$, 2) the survey area coverage and pixelization, and 3) the number of parameters being jointly estimated. Different contaminants will in general have different noise levels, and being able to empirically estimate the noise in the systematics parameters is crucial.

\subsection{Selection Effects}
In both mock catalogs and real data, we need to account for selection effects that systematically modulate the observed number density. For example, these effects can be caused by observing conditions and survey design. Different completeness levels across the survey area can bias the selection of observed galaxies.  %\rachel{That is not a selection bias, however - a selection bias is caused by some effect due to our observation, rather than due to intrinsic population variations}\federico{Addressed, removed incorrect example, only kept one above comment} %\rachel{Need to clarify this term, as it could mean a great many things (e.g., selection in redshift, in area, etc.)}\rachel{Why are we specifying this is an issue for mocks.  I believe the point may be more that it's an issue in real data and in the mocks, right?}\federico{Addressed Both} 
We account for these effects using random catalogs to normalize the observed galaxy number density in every pixel.
\be
\nob_{g}\!(\dir) \rightarrow f_{\text{selec}} \!(\dir)\nob_{g}\!(\dir); \; \; \; 
f_{\text{selec}}\!(\dir) =\frac{N^{\text{tot}}_{g,\text{rand}}}{N^{\text{tot}}_g } \frac{1}{\nob_{g,\text{rand}}\!(\dir)},
\ee
where $N^{\text{tot}}_g$ is the total number of galaxies in our sample and $N^{\text{tot}}_{g,\text{rand}}$ is the total number of random galaxies. Their ratio serves as a normalizing factor given that our random catalogs have higher number density than our survey catalogs. %\rachel{Something seems off in the math here.  If we were to use 100 times as many random points as real galaxies, then it seems to me that both of the factors in that equation will have a factor of 1/100 in it.  Should the first fraction be inverted?}\federico{Yes good catch!} 
In addition, we remove pixels in which our random catalog has very few galaxies to avoid extreme outliers and division by zero. Specifically, we calculate the maximum number of galaxies in a pixel for our random catalog, and discard all pixels that contain fewer than 10\% of this quantity. In practice this resulted in removing less than $1 \%$ of the total area of the mocks used, which are described in Sec. \ref{data and software: mocks}. %\rachel{Say in practice what fraction of the total area and catalog size is removed by this cut, so the practical implications are clear.}\federico{Addressed} 
Note that since we carry out this process using the random catalogs, the results simply remove any survey regions that fall into pixels with very little survey coverage.

%% file: Data_and_Software.tex
\section{Data, Software, and Analysis} \label{data and software}

This section describes the data and software choices used to test the methodology described in the previous section. We start by describing the mock extragalactic galaxy catalog used and the process of generating mock systematic template maps. Following this we explain the software used to fit for the systematics model using the one-point function and estimate the two-point function. Finally we describe the process of quantifying cosmological impact using the data and tools in this section.

%\rachel{Section needs a few-sentence intro.}\federico{Addressed}

\subsection{Mock Extragalactic Galaxy Catalog}\label{data and software: mocks}

To validate our methodology we use the KiDS-like Halo Occupation Distribution (HOD) %\rachel{Have you defined this acronym?}\federico{Addressed} 
$N$-body simulations (KiDS-HOD hereafter) produced by \cite{Harnois2018}. This set of galaxy mock catalogs is produced from the wider set of 
 SLICS (Scinet LIght Cone Simulations) described in \cite{Harnois2015}. The 1025 $N$-body simulations used in SLICS are produced by the gravity solver CUBEP$^3$M \citep{Deraps_2013}, where each run follows $1536^3$  %\rachel{I don't believe this is correct - in $N$-body sims, the particle number is basically always in the form $X^3$ for some value of $X$. And modern sims have orders of magnitude more than a few thousand particles.  Please check/confirm.  My hunch is this should be $1536^3$ rather than $15363$, but that's only a guess based on what is typical for sims.}\federico{Addressed: Typo missed the  symbol} 
 particles inside a grid cube of comoving side length $L_{\text{box}} = 505 \; h^{-1}  $ Mpc. The $N$-body simulation produces a full light cone reconstruction, which requires simulating the non-linear evolution of the particles from a desired redshift, as well as generating halo catalogues. The generated halo catalogs are then used, along with a given HOD model, to assign a certain galaxy population to every host halo. The HOD model used to produce these mocks is described in \citet{Smith_2017} and it is referred to as the GAMA HOD model in \citet{Harnois2018}. For the KiDS-HOD galaxy mocks, the HOD model is calibrated to reproduce key properties of the KiDS Survey \citep{Hildebrandt_2016}. The galaxies are assigned up to redshift $z_{\text{spec}} = 2.0$ and contain photometric redshifts and lensing information as well.

Note that we chose to implement our method directly on the two-point function (in configuration space), which is why we chose to work directly with galaxy positions, rather than looking at contamination in the power spectra. %\rachel{Is this actually more common?  There are many real-space clustering analyses, and I don't think they focus on contamination in power spectra.  I'm not sure the editorializing about `more common route' is helpful.}\federico{Addressed} 
We used a set of 119 publicly-available KiDS-HOD mock catalogs\footnote{See available SLICS products at: \url{http://cuillin.roe.ac.uk/~jharno/SLICS/}} with total sky coverage of $100 \; \text{deg}^2$, along with an associated random catalog. The random catalog has a density approximately 10 times higher than that of the mocks. We work with only one low redshift tomographic bin ($0.2 < z_{\text{spec}} < 0.4$) as a proof of concept for the methodology presented in this work. Table~\ref{tab:mocks_description} summarizes key numbers for the mocks used for this work. The catalogs are pixelized with NSIDE = $512$ (pixel area $\sim 49\;  \text{arcmin}^{2}$), meaning that a typical pixel has $\sim 40$ galaxies.

\subsection{Mock Systematic Maps}\label{data and software: mock_systematics}

%\rachel{The next sentence is an example of a sentence that is true, but that is not that useful in a paper, because it is missing the `why'.  It says what is our guiding principle, but not {\em why} that is our guiding principle, which is important to build confidence in our results.  Please reframe.}\federico{Addressed} 
In order to develop and test our method in a realistic scenario, it is important to generate systematics template maps that emulate the structure and distribution of systematics encountered in real data that will bias the observed number density. To do so, we use the Point Spread Function (PSF) area %\rachel{define acronym the first time it is used}\federico{Addressed} area 
size (a proxy for seeing) as our test case, given that this is a well known observational parameter that can bias the observed galaxy number density depending on how the detection process is carried out, especially for ground-based surveys where regions with worse seeing can exhibit more severe challenges in deblending overlapping galaxy light profiles. %\rachel{Say why this is a useful example.}\federico{Addressed} 
We start by measuring the PSF area size autocorrelation function from HSC Public Data Release 1 \citep{Aihara_2017, Aihara_2017_b}.  %\rachel{define acronym, give reference to HSC survey paper and to PDR1 paper}\federico{Addressed} 
%\rachel{need to define the acronym HSC}\federico{Didn't define it here because I assume I will define it in the Introduction. Will define now and change later in neccesary} %\rachel{It is not clear what this phrase means.  The PSF has nothing to do with redshift, so it is not clear what it means to measure the PSF size autocorrelation function in some tomographic bin.}\federico{Addressed: You are correct, I got confused with the galaxies selected in the mocks} 
%\rachel{I think you are missing a step here, at some point you must be going from seeing size to its overdensity, right?}\federico{Addressed} 
From a pixelized map of seeing size,  with NSIDE = 512, we can construct a map of the seeing size overdensity, which is what we use to calculate the autocorrelation function using the estimator described in Sec.~\ref{data and software: gal_clustering}.  We then fit the correlation function to a declining exponential, as it fits the data quite well in the angular scales between 1 and 500 arcmin, which fall within the scales of interest for measuring clustering. %The fitted exponential is forced to approach zero as $\theta \rightarrow 2\pi$ since we expect the correlations to vanish at large scales. 
%% Not clear the above needs to be said and it sounds like we are doing something to force it (which isn't the case, it's just a feature of the exponential.)
%\rachel{I know it's a fitting function, but there is a physical reason this makes sense, so you should give that reason.  (Rule of thumb: when stating analysis choices, do what you can to avoid them seeming too arbitrary.)}\federico{Addressed} 
With this fit and the use of \texttt{CAMB} \citep{CAMB}, we go from the correlation function to the power spectrum ($w^{tt} \rightarrow C_{\ell}^{tt}$),  %\rachel{Here you start using $t$ presumably for template, but I think you need to say explicitly that this is the connection you are making}\federico{Addressed}), 
where $t$ here refers to the seeing size overdensity template we have constructed from the data. We then fit the power spectrum to an exponential function, as it once again provides a good visual fit to the data in the scales we are interested for clustering ($ 40 <  \ell < 1500$), %.  \rachel{Again, why?  Can you say anything empirically about whether this is approximately a good fit?  It doesn't have to be amazing but there should be some statement about how well this does}\federico{Not fully addressed. Back in the day I fitted to an exponential because it was a good visual fit, do you mean to provide some fitting criterion like "it resulted in the lowest chi-squared from all forms tried"?} \rachel{As a rule when saying something is a good fit, even if you say so qualitatively, you should say on what scales, because that should be connected to what scales you use for the analysis.  So you could leave it qualitative if you want, but you need to say something about the scales.}\federico{Addressed: Both right above and in the sentence talking about the angular correlation I mention the scales in which the fit does well}, 
and find the best fit to be $C_{\ell}^{tt} \propto \exp[-\ell/6]$. However, since the scales probed by HSC are much larger than those we measure in the KiDS-HOD mocks, we adjust the scale of the exponential decay to $1/500$, to emulate the overall behaviour of the correlation function in the range of scales for this analysis.  Note that for this validation test, we are not looking to closely fit for the seeing power spectrum, but rather to have an approximate idea of its spatial correlations as a function of scale. We can then use \texttt{HEALPY} \citep{Healpy} %\rachel{citation} 
to generate full-sky template overdensity maps with NSIDE = 512, and only use the pixels in the same area as our KiDS catalogs.

\begin{figure}
  \center

  \includegraphics[width=0.5\textwidth]{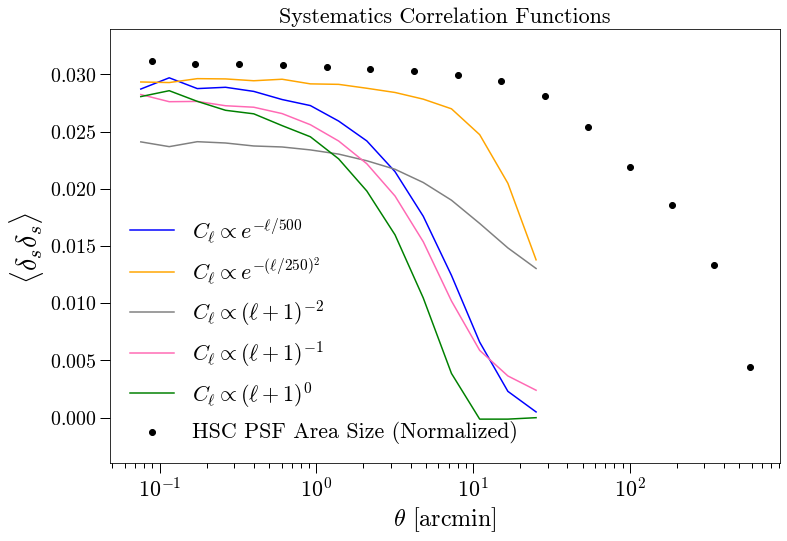}
  \caption{Example of the correlation function for systematics generated through synthetic map-making (solid lines) using one of the 5 systematics families (different colors) described in Subsection~\ref{data and software: mock_systematics} and for the PSF area size residuals in HSC (black points). For visualization purposes, the correlation functions are normalized by their value at $\theta = 1\ \text{arcmin}$. Note that the solid lines are measured on scales appropriate for the analysis done using KiDS-like mocks, which has considerably less area than HSC, resulting in the elimination of larger scales. As we can see, the overall form of the two-point function from our synthetic maps varies amongst different systematic families, albeit with some similarities between certain families. Comparing the solid lines with the real data, we see that the overall exponential behavior of the correlation function and flattening at small scales is similar amongst the systematic families and real data, with the obvious difference that the exponential decay occurs at different scales by design.
  %% RM: repetitive?
  %We see that the systematic correlation functions used in validation is non-trivial and captures some of the expected complexity and behavior we expect from real systematics.
  } 
  \label{fig: syst_correlations}
  
\end{figure}

The above description was for the case of a single systematic. However, we will be interested in generating multiple systematics from different power spectra to introduce a realistic level of variety in the spatial correlations of the systematics. We use 5 different power spectra to represent 5 ``families'' of systematics, where some of the forms are similar to those used in \cite{Huterer}: 

\begin{enumerate}
    \item $C_{\ell}^{tt} \propto e^{-\ell/500}$
    \item $C_{\ell}^{tt} \propto e^{-(\ell/250)^2}$
    \item $C_{\ell}^{tt} \propto (\ell+1)^{-2}$
    \item $C_{\ell}^{tt} \propto (\ell+1)^{-1}$
    \item $C_{\ell}^{tt} \propto (\ell+1)^0$
\end{enumerate}
%\federico{Descirption of Fig.2 starts here}
Fig.~\ref{fig: syst_correlations} shows the correlation function for synthetic systematic maps generated using the above power spectra, as well as the HSC PSF area size correlation function, calculated from HSC PDR1, for comparison. Note that the full-sky maps generated using these power spectra are cut to match the area of the KiDS-HOD mocks, resulting in smaller available scales than those measured for HSC. We see that the overall shape of the correlation function differs for each systematic family, while still being consistent with the expected behavior we see from real data. For example, the flattening of the correlation function at smaller scales and exponential decay at larger scales is seen both the real data and in the synthetic maps. The actual scale on which the correlation function decline differs in the HSC data and the mocks by design, as described above. 

This figure shows that despite the limited area coverage of the mock catalogs that drove us to use smaller scales than we would in a realistic analysis, the shape of systematic correlations across the scales used is non-trivial and depicts the complexity we can expect for real systematics. Maps can then be generated using a choice from these families, and in practice all maps are normalized to have equal variance. We can see that for a fixed variance, there will be different patterns in the   spatial correlations for these families of power spectra. For example, family~(iii) has less power for higher $\ell$ modes or smaller spatial scales than most of the other families, while family (v) can be identified as white noise.

\subsection{Measuring Galaxy Clustering and Contamination Parameters}\label{data and software: gal_clustering}

We estimate the galaxy two-point correlation function using the Landy \& Szalay estimator \citep{Landy} implementation in \texttt{TreeCorr}\footnote{\url{https://github.com/rmjarvis/TreeCorr}} \citep{Jarvis}:

\be \label{eq: landy_szalay}
w(\theta) = \frac{DD - 2DR + RR}{RR} ,
\ee
where D and R represent the galaxy and random catalogs respectively, and DD, RR and DR are the number of galaxy pairs. In our specific implementation, the two-point function is calculated in 15 logarithmically separated bins in angle $\theta$ between 0.06 arcmin and 30 arcmin, a choice made given the small area of our mocks. The pixel size ($\sim \!7$ arcmin) used to estimate the contamination parameters is much larger than the smallest scales used to measure clustering here. The pixel size was chosen in the context of the scales used for current survey science using galaxy clustering \citep{Rodriguez_Monroy_2022}. The reason we include scales much smaller than the pixel size in our analysis is due to the limited size of the mock catalogs ($100 \text{ deg}^2$). Capping our analysis to scales commonly used for measuring clustering in surveys ($\sim 5-200$ arcmin) would have strongly limited the range of scales we can use for this analysis, given that larger scales are greatly affected by low-statistics noise. In addition, the large pixel scale does not cause an issue in our tests since this pixel size was self-consistently used both to construct the template maps and to correct for their contamination. For this reason we include smaller scales than the pixel scale in our analysis, and comment on the fidelity of the corrections in Sec.~\ref{results: 1_syst}.

Covariance matrices for the estimated two point functions are calculated using all 119 mocks. That is, the covariance of two angular separation bins is given by
\be \label{eq: cov}
C_{\alpha \beta} = \frac{1}{N_{\text{realizations}}}\sum_k^{N_{\text{realizations}}}(w_k(\theta_{\alpha}) - \bar{w}(\theta_{\alpha}))(w_k(\theta_{\beta}) - \bar{w}(\theta_{\beta})), 
\ee
where $N_{\text{realizations}} = 119$ in our case,  %\rachel{In previous sections you had used $N_\text{realizations}$, not $N_\text{real}$ - should homogenize across sections}\federico{Addressed} \rachel{No, it's still inconsistent - the equation 23 uses `realizations' while the text uses `real'. 
 %I haven't checked other sections to see about consistency with those, but you should use a consistent notation across text and equations in all sections.}\federico{Yes I changed it everywhere else but for some reason missed the most obvious one, directly after the equations.}, 
 the subscripts $\alpha, \beta$ refers to the angular separation bins $\theta_{\alpha}$ and $\theta_{\beta}$, $w_k$ is the two-point function of the $k$th mock catalog, and $\bar{w}(\theta_{\alpha})$ is the mean correlation function of bin $\alpha$ over all mock catalog realizations.

\begin{figure*}
  \center

  \includegraphics[width=0.6\textwidth]{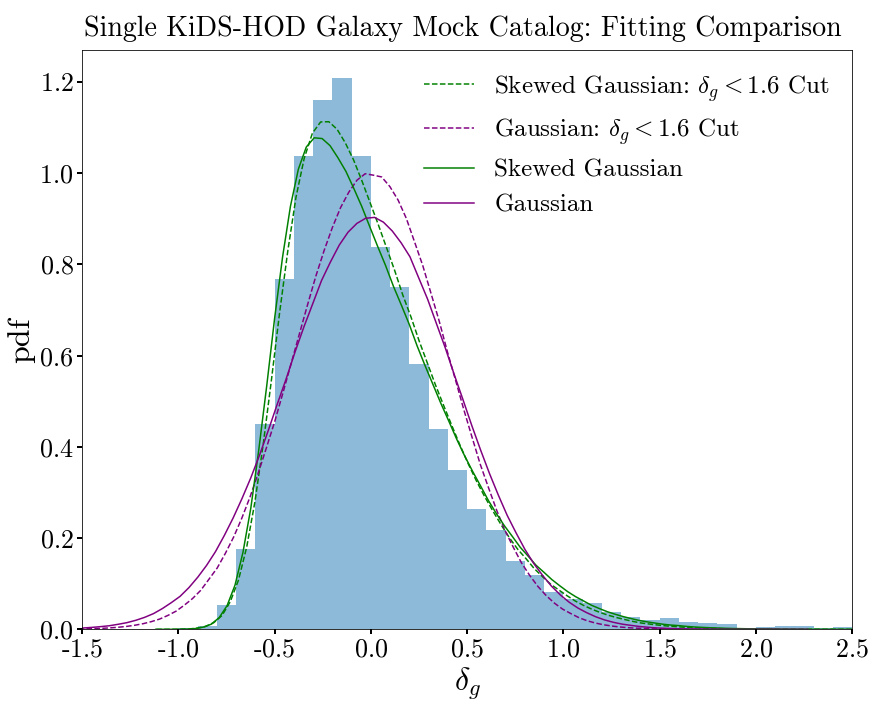}
  \caption{Filled histogram shows the distribution of true galaxy overdensity $\delta_g$ for one systematics-free mock catalog from the KiDS-HOD galaxy simulation. The solid lines show the results of maximum likelihood estimation fits for the distribution with a skewed Gaussian (green) or Gaussian (purple) likelihood. All contamination parameters are fixed to zero while fitting, while $\sigma$ is free to vary in both likelihood forms. The parameter $\gamma = 0$ for the Gaussian likelihood but is free to vary for the skewed Gaussian. The dotted lines show the same fit after removing pixels with $\delta_g \geq 1.6$ as potential outliers.}

  \label{fig: 0syst}
  
  %\rachel{If you make the right panel use lines instead of filled histograms, you could use the same line styles as on the left.  I personally think this would be better, because on the left, the filled histogram represents the ground truth, and the lines represent fits to the observed $\delta_g$ values.  On the right, the filled histograms now indicate fits rather than truth.  It's a bit confusing/inconsistent, and it would be better to have a consistent style.  As a bonus, if you homogenize, then you only need a legend on one of them.  Also, the filled histogram needs to be described in both the legend and the caption.}}
\end{figure*}

We estimate the contamination parameters described in equation~\eqref{eq: mult_syst} %\rachel{Too vague reference - point to a subsection or equation for maximum utility to readers}\federico{Addressed} 
at the level of the one-point function (i.e., the PDF of $\delta_g$ values) by sampling the parameter space using a Monte Carlo Markov Chain (MCMC) algorithm with \texttt{emcee} \citep{Emcee} and choosing the parameters corresponding to the maximum likelihood point. Even though parameters can be more efficiently obtained by using an optimizer (such as gradient descent), we use MCMC to obtain estimates of the parameter noise from single mock realizations (without the need to assume a Gaussian likelihood) and compare them to those from multiple mock realizations. The algorithm needs a user-specified number of walkers to explore the parameter space for a certain number of iterations. The number of walkers should generally be at least twice the number of parameters, while the number of iterations needed to converge might depend on the number of pixels used. For this work, we used $250$ walkers and 1500 iterations. We evaluate convergence heuristically by running the MCMC for 5000 iterations, on a single mock realization, using the maximum number of parameters considered in this work. We confirm that the mean and variance of each parameter, calculated across all chains, stabilizes  to the 5\% level on average after 1500 iterations. We also confirmed that increasing the number of walkers or iterations did not considerably decrease the integrated autocorrelation time of the MCMC chain, which was approximately 90 for a 52 parameter inference. %\rachel{In the caption to figure 1, there is a reference to 100 walkers.  Should reconcile these.}\federico{Addressed: figure 1 should say 250 walkers not 1000 walkers, 1000 was the number of iterations}

%\rachel{Need to comment on convergence tests that gave you confidence the fits had converged.}\federico{Addressed}

\begin{figure*}
  \center
  \includegraphics[width=0.8\textwidth]{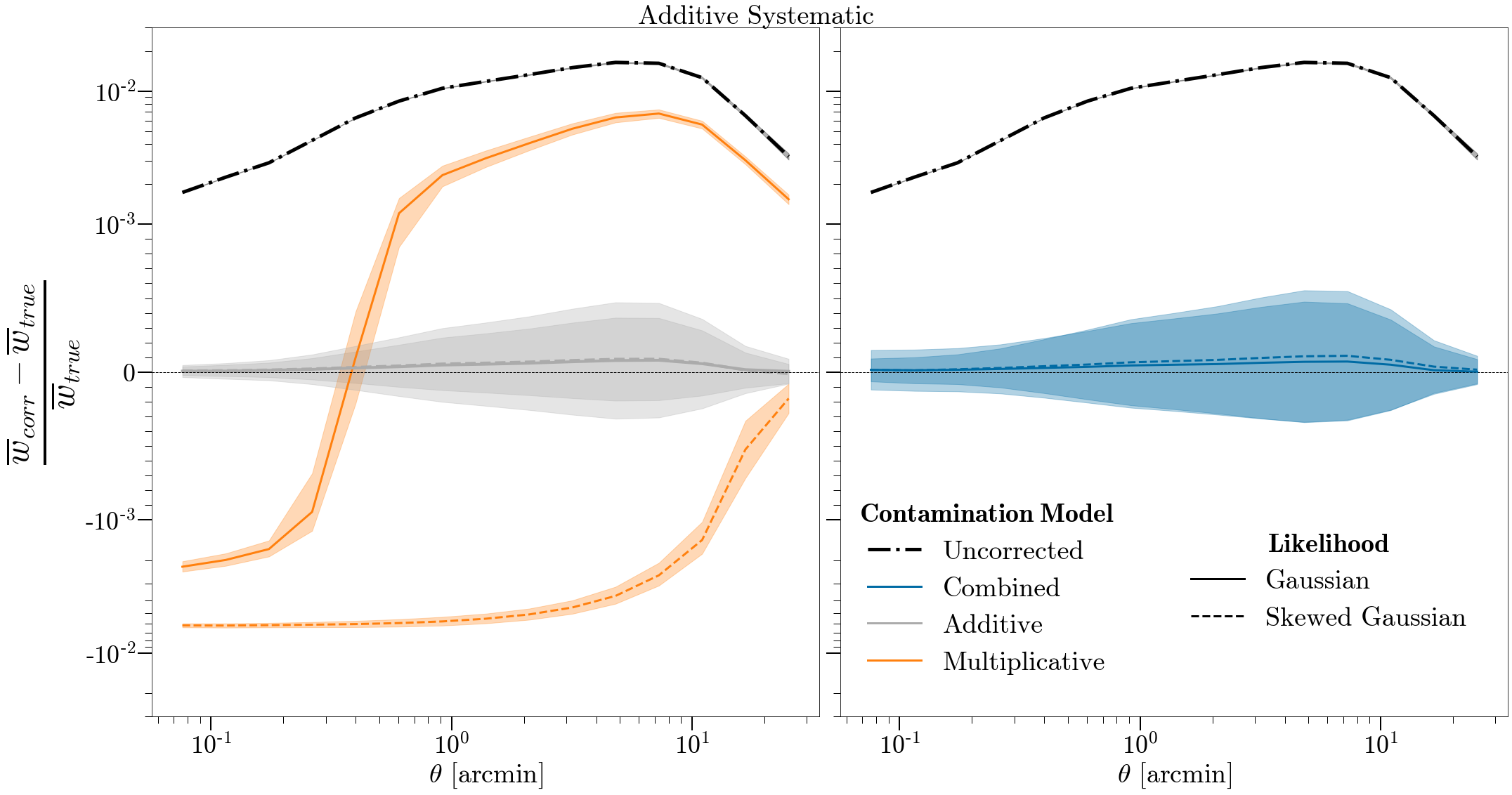}
  \includegraphics[width=0.8\textwidth]{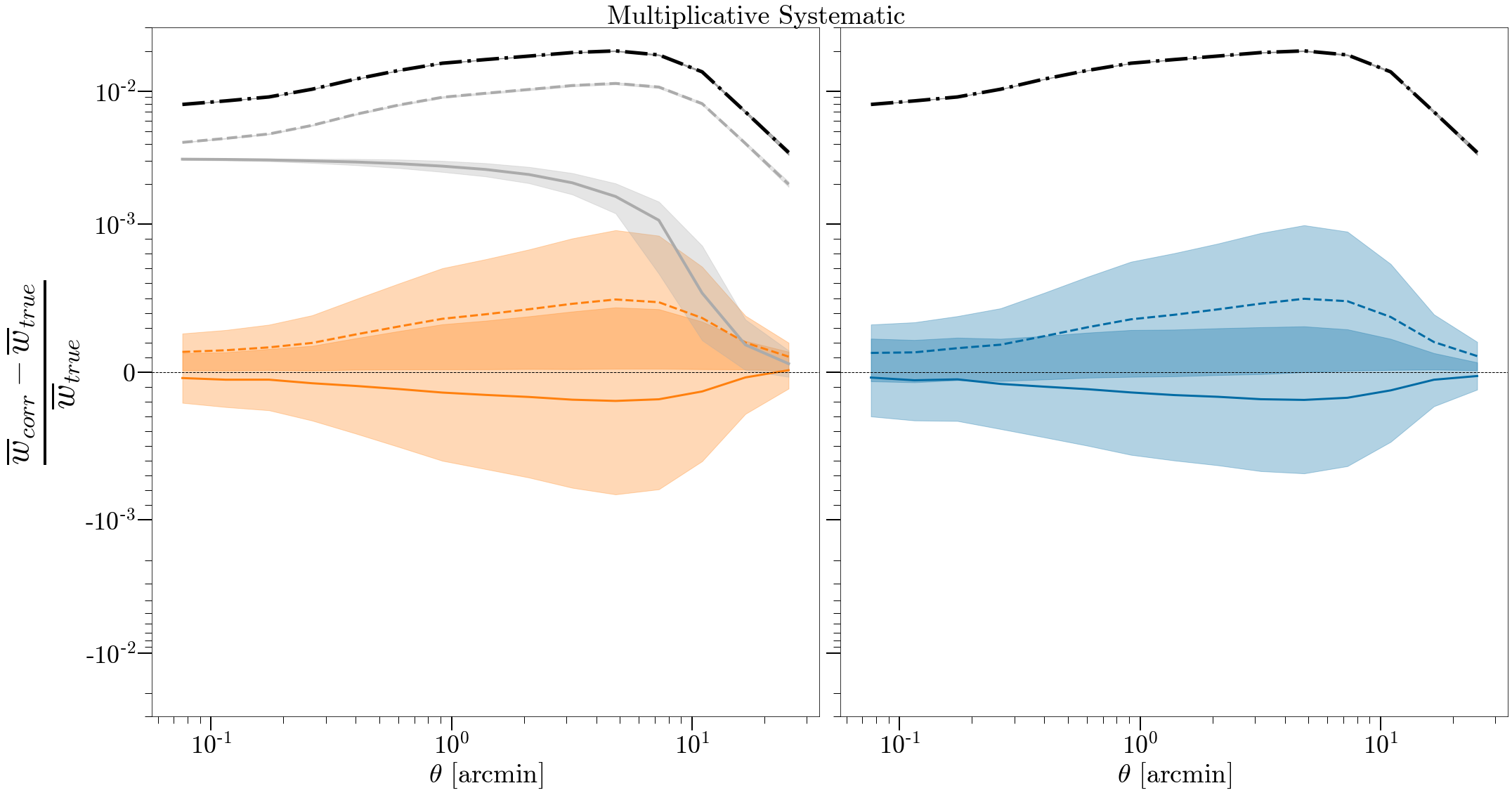}
  \caption{Fractional difference between the corrected and true two-point correlation functions in the presence of a single additive (top row) or multiplicative (bottom row) %\rachel{these labels are opposite from the plot, which says the top is additive and bottom multiplicative -- please check/fix}\federico{Addressed} 
  systematic contaminant. Right and left panels show the same success metric but are separated for visualization clarity. The black dash-dotted line shows the uncorrected two-point function, with contamination of at most 2 per cent. Solid lines show the results after correction using a Gaussian likelihood for the MLE, while dashed lines show the correction using a skewed Gaussian likelihood. The different colors show different contamination models used: additive (grey, left panels), multiplicative (orange, left panels), and combined (blue, right panels). The shaded contours show the $\pm1\sigma$ regions for their respective curves based on multiple realizations of mock catalogs. In both rows, we see that using the incorrect contamination model leads to residual biases after the systematics correction, while correction with the combined model results in equivalent performance to using the true contamination model.  For both the true and combined model, the residual  systematics are below the 0.1 per cent level. There is no significant difference in performance between the skewed and Gaussian likelihoods when using the combined systematics model.}
  \label{fig: 1syst}
  
  %\rachel{You need to explain the significance of left/right panels, and also explain the y-axis style.}}
\end{figure*}

\subsection{Quantifying Cosmological Impact} \label{data and software: clustering_fitting}

We quantify the possible cosmological impact of residual  contamination in the clustering signal by modeling the corrected estimate of the clustering signal as a function of the true signal modulated by an amplitude $A$:

\be \label{eq: corr_bias}
w_{\text{corr}}(\theta) = A w_\text{true}(\theta) \equiv (1+\Delta A) w_\text{true}(\theta). 
\ee
where $ \Delta A = A - 1$ quantifies the bias in the corrected clustering signal. %\rachel{To me, it seems confusing to define $\Delta A$ in terms of a quantity $A$ that you only use for the first time in an equation much later. 
% I suggest that it would help the reader if you are explicit in the equation, like $w_\text{corr}(\theta) = A w_\text{true}(\theta) \equiv (1+\Delta A) w_\text{true}(\theta)$.  You'll notice that I am recommending you define $\Delta A$ with the opposite sign as you've currently done.  The reason for this is that it's traditional to talk about and present results for biases such that a positive number means what you measured is too high.  That's the opposite of your current sign convention.  Though I notice that the table has a positive $\Delta\bar{A}$ value for the uncorrected signal, which is too high, and that makes me think that the equation you've given here does not reflect what you did (i.e., you already did what is in my equation)?  Should check/fix this in the equation here and elsewhere.}\federico{Addressed. Yes the sign was flipped in the definition of $\Delta A$, I fixed it.} 
This can be understood in the context of 3$\times$2pt analysis, where clustering and galaxy-galaxy lensing are used to constrain the galaxy bias, as explained in subsection \ref{background: cosomology}. The most relevant dependence in this context is that of $\sigma_8$, a measure of clustering strength, which in the linear regime is related to the galaxy bias, $b$, and galaxy clustering, $w_{gg}$, by the relationship: $w_{gg}\propto b^2 \sigma_8^2$. Measuring any biases on the amplitude of $w_{gg}$ will determine how accurately the $b\sigma_8$ combination can be measured. %\rachel{I don't find this explanation very satisfying.  You haven't said anything about galaxy bias, which is clearly there.  My recommendation is to make this discussion more specific.  For example, in the context of 3x2pt analysis, clustering and lensing together are used to constrain the galaxy bias.  In that context, probably the most relevant dependencies are that $w_{gg}\propto b^2 \sigma_8^2$ in the linear regime.  So, biases on the amplitude of $w_{gg}$ will determine how accurately the $b\sigma_8$ combination can be measured. Making it more specifically about this will be a cleaner narrative than ignoring the galaxy bias without explanation.}\federico{Addressed} 
 In our validation case of $119$ mocks, the value of A can be obtained for every individual mock catalog using the corrected and true estimates of the two-point function by minimizing the $\chi^2$:

\be \label{eq: chi_2}
\chi^2 = (w_{\text{corr}} - Aw_{\text{true}})^T \; \bm{\mathsf{C}}_{\text{corr}}^{-1} (w_{\text{corr}} - Aw_{\text{true}}),
\ee
%\rachel{Need to be more specific that this is being done using the estimate of the true and corrected correlation function for a specific mock catalog, while the covariance is across all mocks.  The equation is only schematic and lacks the precision of previous equations that were very clear about whether they apply to a specific realization or across realizations.} \rachel{Also the text is a little vague - are you actually doing a fitting process?  I think you don't need to - the MLE for $A$ must be something like $(w_\text{true}^T C_{\text{corr}}^{-1} w_\text{corr}) / (w_\text{true}^T C_{\text{corr}}^{-1} w_\text{true})$, right?}\federico{Addressed: Correct you don't need to do a fit technically, I've changed the wording to just say "minimize the chi-squared". Also clarified this is for every mock prior to the equation.}. 
We use the full range of scales from the measured clustering (0.06  -- 30 arcmin) in this fitting process. Doing this fit for every mock catalog, we can compute the average over all realizations, $\bar{A}$, to obtain a measure of the bias. %\rachel{Should probably give this a symbol, like $\bar{A}$ or $\langle A\rangle$, to distinguish it from the specific quantity in equation 25 measured for one realization.  And then use that new symbol in the rest of the discussion (since you often mean the average rather than the outcome for one realization)}\federico{Addressed}. 
The covariance matrix $\bm{\mathsf{C}}_{\text{corr}}$ here is taken from all realizations of the corrected clustering signal and is calculated using equation~\eqref{eq: cov}. %\rachel{point back to your equation that says how to get this}\federico{Addressed}. 
%\rachel{Do we need this complex symbol or can we just call it $\sigma_A$ throughout this section?}\federico{Because its the standard error and not the standard deviation, I wonder if some distinction with just using $\sigma$ is useful or not.} 
 We are interested in the estimated bias on the corrected two-point function, $\Delta \bar{A} = \bar{A} - 1$, as the success metric to quantify the significance of residual bias in the corrected two-point function. Note that using the realization-specific $w_{\text{true}}$ and $w_{\text{corr}}$ to estimate $A$ for every mock catalog largely removes the impact of cosmic variance on $A$, reducing the uncertainties in our analysis. %Meaning that a measure of the standard deviation of all measured values of $A$ should be free of any cosmic variance. 

%\rachel{Should comment specifically on what the above process means for noise in the process - i.e., you've done this in a way that essentially removes the impact of cosmic variance.}\federico{Addressed}

\subsection{Summary: Step-by-Step Procedure}\label{data and software: procedure}
We summarize the previous  subsections with a step-by-step procedure for generating and analyzing the data of a single mock catalog.

\begin{enumerate}
    \item Measure the true correlation function of the extragalactic galaxy catalog described in Sec.~\ref{data and software: mocks} and generate pixelized maps with $\text{NSIDE} = 512$.
    
    \item Generate $N_{\text{sys}}$ full-sky template maps using \texttt{HEALPY} from the desired selection of the 5 systematics families described in Sec.~\ref{data and software: mock_systematics}. Mask the full-sky map to only cover the same area as the galaxy mock catalog. Compute the correlation function for all systematics by assigning each galaxy the corresponding template values for the pixel in which the galaxy is located. 

    \item From the pixelized versions of the galaxy field ($\delta_g$) and systematics maps ($\delta_t$), we generate the contaminated galaxy field ($\dob_g$) using equation ~\eqref{eq: mult_syst} from a chosen set of contamination parameters. We then calculate the contaminated clustering signal using equation ~\eqref{eq: obs_w}.

    \item With the observed galaxy overdensity field and systematics maps as inputs, the contamination parameters are jointly estimated as prescribed in Sec.~\ref{data and software: gal_clustering} by using a MCMC algorithm with the chosen likelihood model (Gaussian or skewed Gaussian) and contamination model (additive, multiplicative, or combined). 

    \item The contaminated clustering signal is corrected using equation~\eqref{eq: corr_w_uncorrelated}.
\end{enumerate}
This procedure is followed for all 119 mock realizations, yielding an ensemble of results used to quantify decontamination quality, parameter estimation and noise, and cosmological impact.

%% file: Results.tex
\section{Results} \label{results}

\begin{figure*}
  \center
  \includegraphics[width=0.75\textwidth]{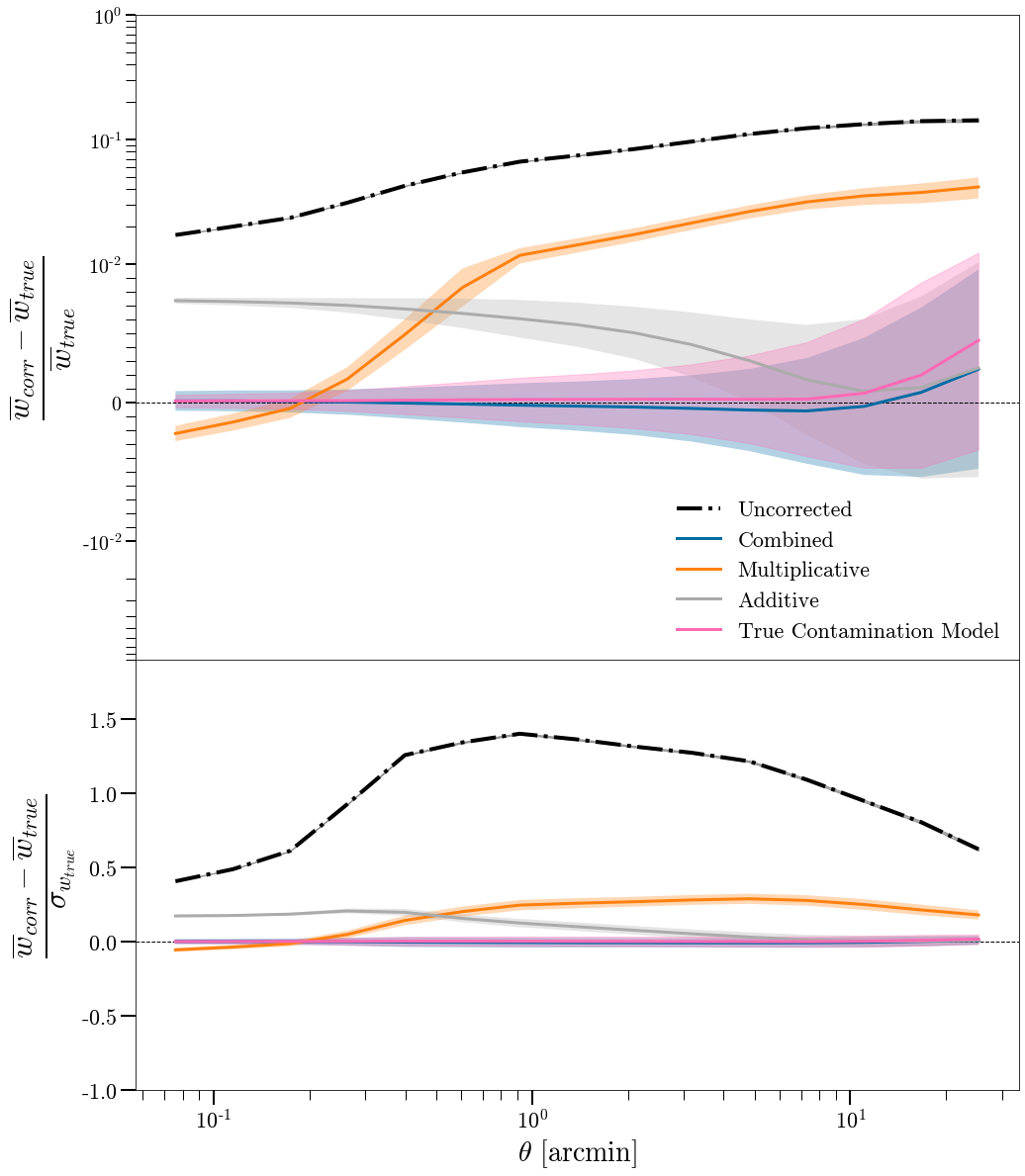}
  
  \caption{Top panel: Fractional difference between the average corrected and true two-point function in the presence of 25 systematic contaminants, 10 of which are additive and 15 multiplicative. The vertical symlog scale shows cases of possible over/under correction.  The black dash-dotted line shows the uncorrected two-point function, with contamination reaching a maximum of $\sim$10 per cent. Solid lines show the estimated correction using a Gaussian likelihood for the MLE and the combined (blue), additive (grey), multiplicative (orange) or true (pink) contamination model. %\rachel{Is there a reason you aren't mentioning the above models?  Also, orange seems to be multiplicative, not true.}\federico{Addressed} 
  The shaded contours show the $1\sigma$ regions of their respective lines. Bottom panel: The difference between the corrected and true two-point function divided by the standard deviation of the true clustering across all realizations for each separation bin. Both panels show that the combined model robustly mitigates contamination to the sub-percent level with respect to the clustering signal and well within $1\sigma$ on all scales, and on average is equivalent to using the true contamination model. The performance of the true and combined models is consistent within the statistical uncertainties. %\rachel{Should comment explicitly on the vertical symlog scale since it's non-standard.}\federico{Addressed at begining of caption} 
  }
  \label{fig: 25syst_correction}
\end{figure*}

In this section, we show the results of applying our method to mock galaxy catalogs.  We will first apply our method to mock catalogs without any  systematics, to explore differences between our use of a Gaussian or skewed Gaussian distribution for $\delta_g$ (Sec.~\ref{results: 0_syst}). We will then test our systematics mitigation method in different scenarios varying in complexity to evaluate its performance. The outcome of that exercise, which will be detailed in subsections below, may be summarized briefly as follows:
  
\begin{itemize}
    \item Sec.~\ref{results: 1_syst} shows that systematics mitigation performance is especially sensitive to choice of systematic model.
    \item The use of a nested contamination model (here defined as "Combined") allows flexibility in defining a model that can describe the multi-systematic case in Sec.~\ref{results: mult_syst}.
    %% RM: I think referencing a single figure breaks the flow and you aren't conveying an actual result here.
    %\item Fig.~\ref{fig: 25syst_cornerplot} demonstrates how accurate parameter and noise estimation can help us understand how individual systematics contaminate the number density (e.g. additive, multiplicative, or other) as seen in. 
    \item %\rachel{This is not a sentence - need to fix:}\federico{Addressed} \rachel{No, it was still not a sentence (there's no subject). I just edited to fix it. }
    Sec.~\ref{results: cosmo_impact} shows that our method can robustly mitigate systematic contamination in the clustering signal to the sub-percent level, reducing the bias on the two-point function to below $0.1\sigma$. %\rachel{I think we have agreed that people will care as much if not more about the systematics mitigation expressed in terms of number of $\sigma$, so I suggest giving that number here as well.}\federico{Addressed}
\end{itemize}

%\rachel{In the preamble to a section, if you have a bulleted list like that, you should try to tie it explicitly to the subsections (otherwise you aren't really giving the reader a road map).}\federico{Do you mean to reference the subsections explicitly? Or are you saying that the conclusions aren't tied to the subsections?} \rachel{I meant the former.  Doing so will help you determine whether you might be missing any key outcomes, or whether your outcomes are not cleanly reflected in the section text.}\federico{Addressed. The middle bullet point has a reference to a paritcular figure rather than a section, I felt it made more sense than specifying the same subsection as the previous bullet point. Could also just refer to the same subsection again if you think thats better.}

%%% here goes figure 4
\begin{figure*}
  \center
  \includegraphics[width=0.9\textwidth]{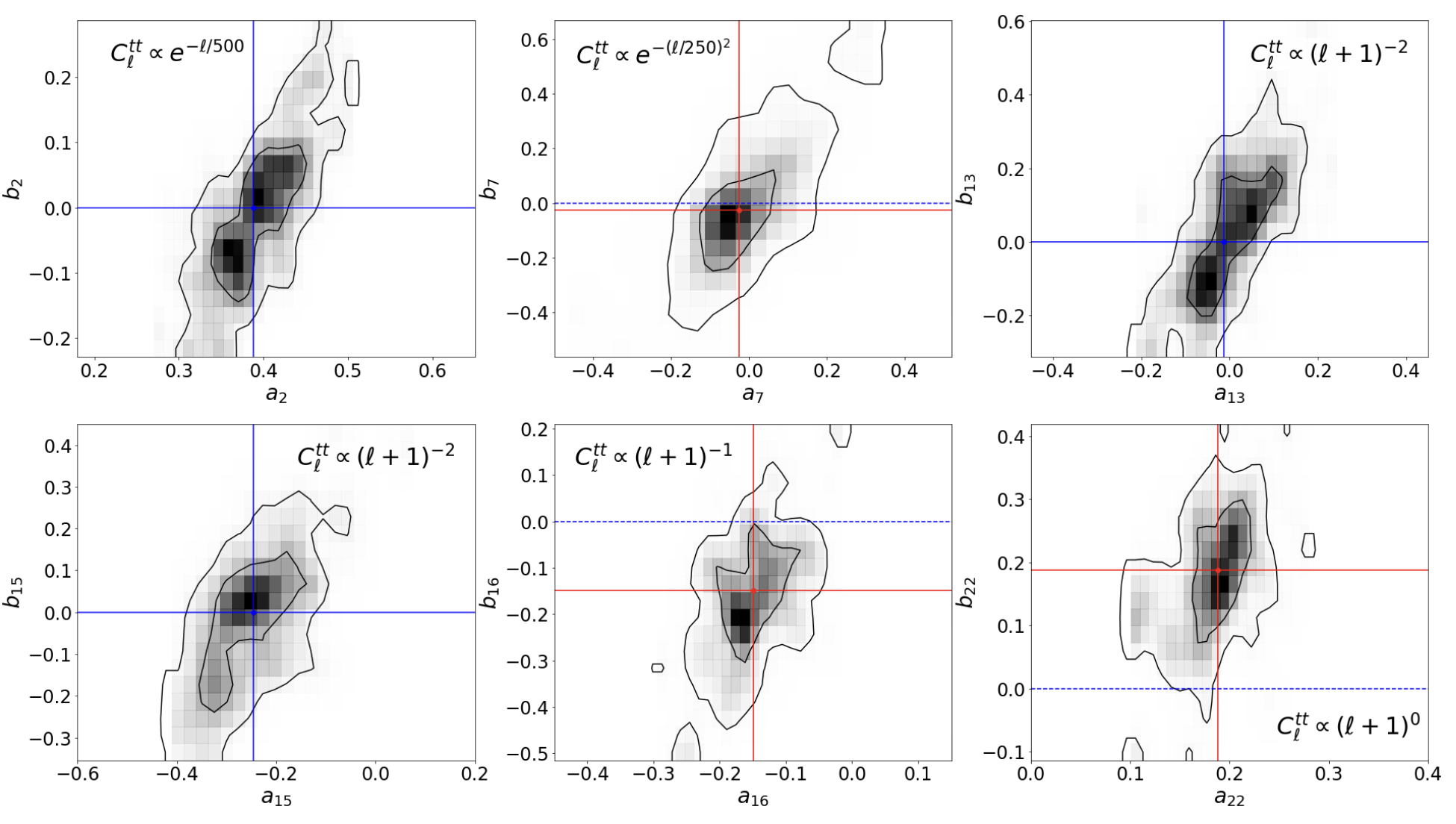}
  
  \caption{Each panel shows a density plot of the estimated contamination parameters, $\{a_i, b_i\}$, using a Gaussian likelihood, from all 119 mock catalog realizations. In other words, every point on a particular panel represents the maximum likelihood estimate from all MCMC samples of a single mock catalog. The contour regions show the $1\sigma$ (68\%) and $2 \sigma$ (95\%) regions. The 6 systematics shown (out of a total of 25 contaminants) form a representative sample of the different systematic families (shown by the scaling of their power spectra, $C_{\ell}^{tt}$) used to generate the systematic maps. The vertical and horizontal solid lines show the true parameters used for contamination, with blue (red) representing  additive (multiplicative) systematics. %\rachel{I'm confused, the text said that the systematics were strictly additive or multiplicative, so I thought one of the truth lines would always be at zero, but that does not seem to be the case.}\federico{I'm not sure I understand. For the blu lines (additive systematics), there should always be a horizontal line at b =0, but for the multiplicative systematics (red lines) neither line has to be at zero. I don't see why the truth lines would always have to be at zero for the multiplicative systematics.} 
  The dashed blue line shows the zero point of the $b$ parameter, where an additive systematic should live in parameter space. We see that across all panels the true parameters fall within $1\sigma$ regions of the estimated parameters and we can use the inferred parameters to identify additive and multiplicative systematics. %\rachel{Axis labels are too small.} \rachel{One thing I have been pondering about this figure is that it seems $a$ is easier to constrain than $b$ (tighter constraints on $a$), but the fact that the figures do not have an equal aspect ratio makes this hard to see.  Have you tried forcing them to an equal aspect ratio so this will be more apparent?  You don't want to have too much white space in the figure, so it might not work well, but I think it is worth trying.  It's also worth flagging this conclusion (re: constraints on $a$ vs.\ $b$) and explaining/interpreting it in the paper text.}\federico{Addressed: The axis labels are fixed. I tried the second point you made in this image here, making the aspect ratios the same. I relaxed the ratio condition a bit on the last 2 panels as it started to look a bit too squished, but let me know what you think of the change. If you wish to see what it looks like without the change, modify the name of the image from CornerPlot3 to CornerPlot2. I included an explanation for the constraint differences in the text following the figure being mentioned}
  }
  \label{fig: 25syst_cornerplot}
\end{figure*}

\begin {table} 
\begin{adjustbox}{width=0.8\columnwidth,center}
\begin{tabular}{ ||c|c|c||} 

\hline
Likelihood Form  & \multicolumn{2}{|c|}{\% Correct} \\ 
\hline
 & Additive  & Multiplicative  \\
\hline
Skewed Gaussian & 100\% & 99\%  \\ 
\hline
Gaussian & 97\% & 93\%\\ 
\hline
\end{tabular}
\end{adjustbox}
\caption {Model comparison using likelihood ratio tests for single systematic cases. The columns show in what fraction of all mock catalog realizations the likelihood ratio test preferred the correct (and simpler) model over the combined model.}
\label{tab: lk_ratio}

\end {table}

\subsection{No Systematics}\label{results: 0_syst}

Before investigating systematics  mitigation, we first explore the underlying distribution we are trying to model. The likelihood function in equation~\eqref{eq: likelihood} %\rachel{should refer to an equation}\federico{Addressed}
is based on certain assumptions about the true $\delta_g$ distribution. For the case where we cannot assume $|\delta_g|\ll 1$, we know that the underlying distribution cannot be Gaussian, as $\delta_g$ is mathematically bounded to be greater than $-1$ but can take any positive value. As seen in Fig.~\ref{fig: 0syst}, we use the true $\delta_g$ field from our mock catalogs (filled histogram) and fit its distribution using both a Gaussian and a skewed Gaussian likelihood function. For both likelihoods we fix any contamination parameters \{$a,b$\} to zero and allow $\sigma$ to vary, while we fix $\gamma = 0$ for the Gaussian likelihood and let it vary for the skewed Gaussian. The solid lines in the left-hand panel show that the skewed Gaussian likelihood provides a better fit to the data than the Gaussian likelihood, given the skewed distribution of $\delta_g$ values.  

%The right-hand panel shows the difference %\rachel{What does `an additional difference' mean?  i.e., what is additional about this difference?}\federico{Addressed: additional does not make sense here, rephrased.} 
%in the estimated variance between the models. 

However, we found that this difference was mostly due to potential outliers, which once removed (shown as the dashed lines) provided better fits to the distribution and agreement in the estimated variance. Physically the outliers represent regions of very high galaxy density. The cut $\delta_g < 1.6$ was made to test this. Note that in a real cosmological analysis, larger scales will be used and the variance will therefore be smaller.

It is important to highlight that this cut was made in the true $\delta_g$ distribution, which we do not have in real data.  Therefore, this cut is purely for illustration purposes and is not applied at any point in the analyses later in this section. However, it is important to point out that the fits with the Gaussian distribution were especially sensitive to these large $\delta_g$ values. As we can see from the right panel of Fig.~\ref{fig: 0syst}, the outliers contributed a considerable portion of the overall estimated variance despite being a small fraction ($\sim \! 1\%$) of the pixels. %\rachel{Should say what fraction}\federico{Addressed}. 
In contrast, the skewed Gaussian likelihood  seems to be less sensitive to outliers. This consideration is relevant because real data will have outliers for a variety of reasons (e.g., survey window complexity not represented in these mock catalogs).  Our results indicate that the skewed Gaussian likelihood may be able to more accurately estimate the true variance of the $\delta_g$ distribution and possibly improve estimates of the systematic contamination parameters modulating the variance, more specifically the $b$ parameter in equation~\eqref{eq: gal_overdensity}.

\subsection{One Systematic}\label{results: 1_syst}

%\rachel{One thing we do not comment on explicitly in the results section is that our likelihood model is glaringly wrong in treating the distribution of $\delta_g$ values in each pixel as independent, yet this model mis-specification seems to not matter (since we are using mocks that include realistic clustering).  My suggestion is to add a short additional paragraph at the end of this subsection: `Finally, we note that our mock catalogs include a realistic cosmological clustering signal, meaning the values of galaxy overdensity in each pixel are correlated.  Our results suggest that even though our likelihood model for $\delta_g$ omits these correlations, this model mis-specification does not affect the fidelity of the systematic correction.  The favorable results for our method with multiple systematics, as shown in subsequent subsections, further support this conclusion.'}\federico{ Yes you are right that is not mentioned, I added this excerpt at the end of the section, thanks!}

We next test the method for a scenario with a single source of systematic contamination.  We consider separately the cases of a single additive and a single multiplicative systematic.  The true overdensity map is contaminated using a single systematic template map generated from the power spectrum (i) described in Subsection~\ref{data and software: mock_systematics}  %\rachel{need to refer to a subsection, not an entire section}\federico{Addressd} 
to produce the observed galaxy overdensity map. The observed two-point function is calculated using equation~\eqref{eq: landy_szalay}. %\rachel{Is that really the case?  If you calculate the observed two-point function that way, then it seems like there is no need for a contaminated overdensity map (to use that equation you only need the template and the true overdensity).}\federico{The contaminated overdensity maps is the input distribution in the MLE to estimate the contamination parameters.} 
In each realization, we estimate the contamination using the parameters from the maximum likelihood estimation approach described in Subsection~\ref{methods: likelihood_func} by fitting for a single systematic map,  %\rachel{again, give a subsection rather than an entire section}\federico{Addressed}, 
then correct the two-point function using equation~\eqref{eq: corr_w_uncorrelated}. We use different versions of the likelihood model in equation~\eqref{eq: likelihood} to test different assumptions about the distribution of $\delta_g$ and its contamination. %\rachel{I think you need to say also that we use different versions of the likelihood for these tests, and point to the relevant equation.}\federico{Addressed} 
In Fig.~\ref{fig: 1syst}, we show the fractional difference between the uncorrected, corrected and true clustering signals to evaluate performance. For context, the systematic correlation function is approximately 2 orders of magnitude smaller than the signal.

%The dash-dotted black line shows the uncorrected (or observed) clustering signal, with at most 2 per cent contamination on the scales shown. We compare the corrected results with the  Gaussian likelihood model (solid lines) against those with the skewed Gaussian model (dashed lines), for three assumed systematic contamination models: combined (blue), multiplicative (orange), and additive (grey). 
%\rachel{Personally, I'd suggest making this more of a discussion rather than trying to flag line styles and colors.  If you include line styles/colors here, then any change in those means you have to edit your caption and text.  It's a lot to keep track of.} \federico{Understood. Discussion of this figure follows this}

%%% Table 3 here used to go here

As expected, assuming the wrong model for the systematic contamination can result in large residual biases even after applying our systematics mitigation scheme, while choosing the correct model results in removal of the systematic to below the 0.1 percent level. The combined model performs almost identically to choosing the correct model, without substantially increased uncertainties despite the small increase in model complexity (one additional parameter).  The similarities in performance between the true and combined model suggest that the combined model is a viable alternative for use in real data. The choice of the likelihood form (Gaussian vs.\ skewed Gaussian) seems to have no significant impact on performance, so choosing the simpler Gaussian model is preferred in this case. However, given the generalized contamination model chosen, we felt it was worth revisiting the question of likelihood choice and test its impact directly. The marginal differences between likelihoods found here reinforces previous works' conclusions regarding the Gaussian approximation for systematics mitigation in galaxy clustering \citep{Huterer, Rezaie_2021}.

%% RM: new paragraph, as these are now some smaller details.
As described in Subsection~\ref{methods: noise_debiasing}, we emphasize the importance of debiasing the estimated model parameters prior to estimating the corrected correlation function. We have confirmed that skipping this step results in substantial overcorrection, increasing the size of the residual biases by an order of magnitude in this case. The severity of the degradation will depend on the size of the parameter estimation noise. We also note that the estimated correction is not degraded below the pixel size used to estimate the contamination parameters, confirming our choice to use scales below the pixel size for this validation.  %\rachel{Should say something quantitative here or point to an appendix rather than asserting there are performance degradations without indication of their severity.} 

We used likelihood ratio tests to see whether we can identify any model preference  from the data, focusing on the choice of combined versus the simpler additive and multiplicative models. We compare the use of both distributions in Table~\ref{tab: lk_ratio} for each systematic model. The columns show in what fraction of all mock catalog realizations the likelihood ratio test could not reject the correct (and simpler) model. In practice, the inability to  reject a simpler model can be used as a form of model selection in favor of that model. In general this test correctly identifies the contamination model in the single systematic case; when using a skewed Gaussian likelihood, it correctly identified the  additive or multiplicative model 100\% and 99\% of the time, respectively.  The Gaussian likelihood model resulted in 97\% and 93\% accuracy. Here, too, there are small differences between using either likelihood model, but both produce strong results overall. This is a positive sign that likelihood ratio tests work well for simple cases such as the single systematic contamination and can potentially be used for model selection. In practice we do not know what the true model is, so this tool can be used to compare systematic models. For a single systematic, the fitting can be done using all three contamination models used here (additive, multiplicative, and combined). Since the combined model serves as a nested model, the likelihood ratio for the additive and multiplicative models can be calculated to evaluate if either form is preferred over the other.

Finally, we note that our mock catalogs include a realistic cosmological clustering signal, meaning the values of galaxy overdensity in each pixel are correlated.  Our results suggest that even though our likelihood model for $\delta_g$ omits these correlations, this model mis-specification does not affect the fidelity of the systematic correction.  The favorable results for our method with multiple systematics, as shown in subsequent subsections, further support this conclusion.

%\rachel{I feel like the punch line is not clear, because in practice, you've compared combined vs.\ true model.  But in real life, we don't know the true model.  So how can the test you've described here be used in practice for model selection?}\federico{Addressed: let me know if this makes sense.}

\subsection{Multiple Systematics}\label{results: mult_syst}

%\rachel{One thing that just occurred to me is that we do not actually present tests for a systematic where the true model is neither additive nor multiplicative.  This is a bit unfortunate as it's a case where our method should particularly excel.  My suggestion is that at the end of this subsection we should add something like: `As shown here, the introduction of the combined model allows for joint modeling and correction of a set of clustering systematics that have different functional  forms.  Similarly, by construction it will allow for modeling of systematics that are neither purely additive nor purely multiplicative in overdensity.'}\federico{Yes that sounds good, added.}

We now move to the treatment of multiple systematics. When dealing with real survey data (e.g., DES), a multitude of systematic templates \citep{Rodriguez_Monroy_2022} are considered during the mitigation process, motivating our extention to the case of multiple systematics. We use a non-trivial number of uncorrelated contaminants, 25, to test our method. We generate 5 independent templates from each of the 5 systematic families described in Sec.~\ref{data and software: mock_systematics}, resulting in 25 independent templates.  %\rachel{25 templates from each family would mean 25 templates $\times$ 5 families which gives 125 templates.  But you only have 25 systematics.  So is this what you really mean?  Or do you mean something like `5 independent templates from each of the 5 systematics families, giving 25 independent templates'?}\federico{Addressed}
%\rachel{Isn't it 25 templates from 5 families? 
 %In this section the focus is on uncorrelated systematics, so even for the 5 systematics from the same family, they must be independent realizations, right?}\federico{Addressed: yes should be 25, not 5, it was a typo. Yes even for the same faimily they are independent. } 
 The contamination model used is chosen with a random number generator such that each systematic has equal chance of being additive or multiplicative, resulting in 10 additive and 15 multiplicative systematics for this analysis. The nonzero contamination parameters  %\rachel{Should this be `the nonzero contamination parameters'?  e.g., for the additive case, one of the parameters must be fixed at 0.}\federico{Addressed} 
 are chosen from a Gaussian centered at zero and standard deviation of $0.15$. This particular value of the standard deviation is motivated by real observations of seeing in HSC and chosen to result in an approximate 10\% contamination level at larger scales when using the systematic families described in Subsection \ref{data and software: mock_systematics}. This level of contamination is reasonable considering the results of \cite{Rodriguez_Monroy_2022}. Using these choices we produce the observed galaxy overdensity field for each realization of the mocks using equation~\eqref{eq: mult_syst}. %\rachel{Probably should point to a specific equation, like equation 12 perhaps, so it's clear exactly what was done.}\federico{Addressed} 
 We then fit for the contamination parameters jointly for all systematics using the combined, additive, and multiplicative model, as well as the true contamination model for comparison. To clarify, in the additive and multiplicative case all systematics are assumed to follow that specific model, while in the true contamination model, each systematic follows their true assigned model when fitting.   %\rachel{You may need to emphasize that in the first case, you assume {\em all} systematics follow the same specific model, whereas in the `true' case, you let each them follow different models (either additive or multiplicative).}\federico{Addressed}

We compare performance of the different contamination models in Fig.~\ref{fig: 25syst_correction}. We show only the results using the Gaussian likelihood for the fits, as we found negligible difference when using the skewed Gaussian form. Both panels show that the combined model produces sub-percent mitigation of the contaminated two-point function (to a very small fraction of the statistical uncertainty) and equivalent performance to using the true contamination model within the statistical uncertainties.  On the other hand, using the additive and multiplicative models consistently across all systematics results in biased corrections of the clustering signal (model mis-specification bias). We conclude, therefore, that assuming a single contamination model in the presence of systematics with distinct types of contamination can lead to biased estimates of the correlation function. In this case, the safest choice when the form of systematics is not known is to use the combined model for all of them. We note that the increased degradation at larger scales is due to the small area of the KiDS mocks. This was confirmed using larger simulated mocks described Sec.~\ref{discussion: practical_applications}, where this degradation is not present at these scales. We also verified that the combined model can still recover the truth and outperforms the simpler models in the case of no contamination ($a = b = 0$) for a random selection of templates.

\begin{figure*}
  \center
  \includegraphics[width=0.6\textwidth]{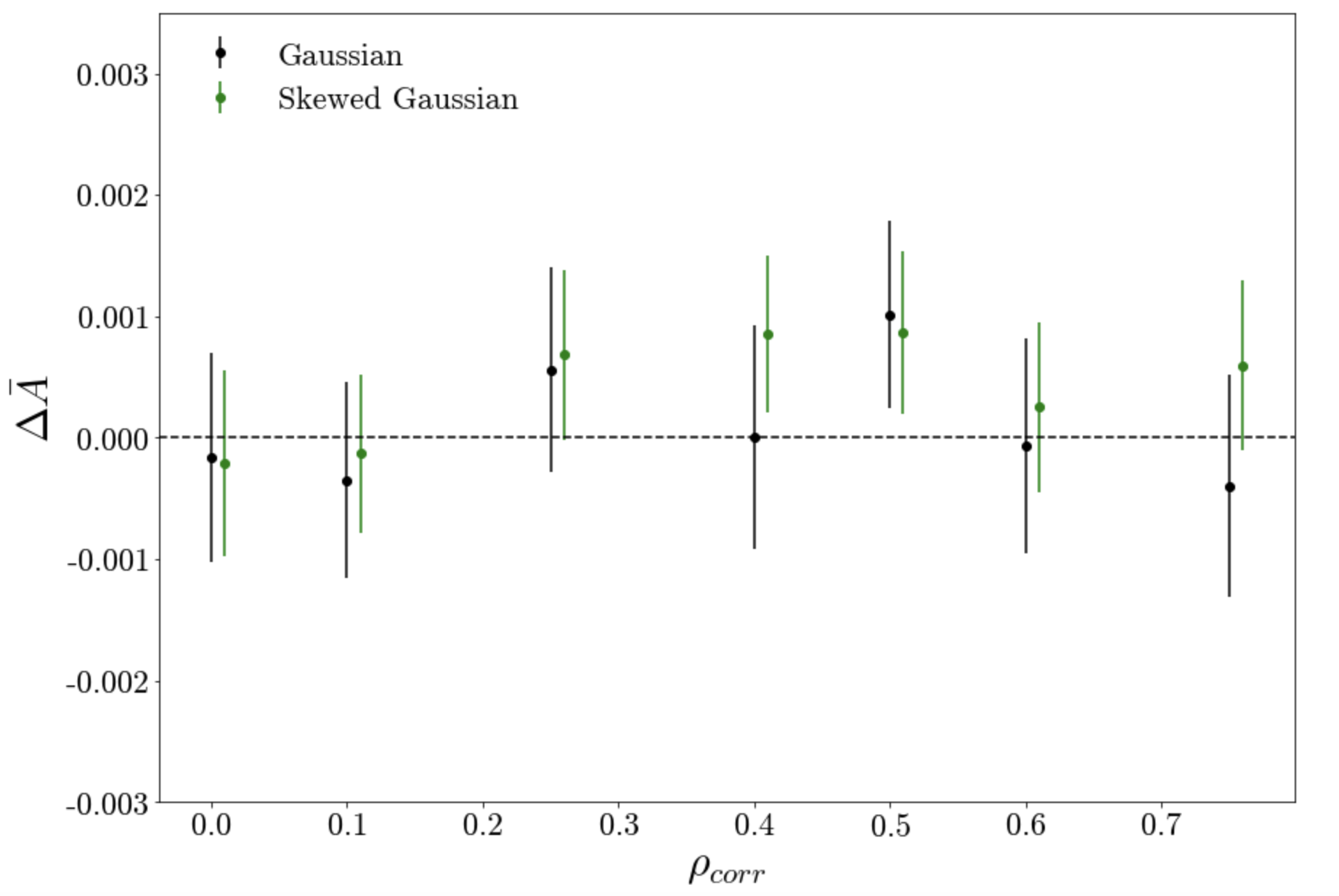}
  
  \caption{The average bias on the corrected  clustering signal over all 119 mock catalog realizations as a function of the correlation coefficient between systematic templates within a given power spectrum class. The solid points show this metric for the corrected two-point function using the combined contamination model and either a Gaussian (black) or skewed Gaussian (green) likelihood in the MLE. The errorbars show the error on the mean. %\rachel{I do not see a grey shaded region.}\federico{Addressed: This part of the caption was for old version of figure, shaded region does not apply to this version of figure} 
  %This metric is calculated for different levels of in-class correlation between systematic maps. 
  This figure shows that the excellent performance of the method remains unaltered even by relatively high levels of correlation between systematic template maps. %\rachel{I don't understand the last sentence.} \federico{Addressed: Again this comment is from an older version of the figure that was incorrect and had huge error bars. I deleted the comment.}
  }
  \label{fig: 25syst_bias}
\end{figure*}

Finally, we consider whether we can learn anything about the form of our contaminants (e.g., are some purely additive or multiplicative?) using our methodology. We avoid using likelihood ratio tests for this, as with multiple contaminants, comparing forms for each one is not as trivial as in the single systematic case. To see if we can learn something about the underlying contamination model we use the estimated parameters directly, i.e., the best-fitting values for $\{a_i, b_i\}$ from using the combined model.  %\rachel{In the example that follows, we need to include statistical uncertainties, because they will be present in reality and are part of the interpretation:}\federico{Addressed} 
For example, if we estimate the contamination parameters and uncertainties for systematic $t_i$ to be: $\{ \hat{a}_i = 0.5\pm 0.03, \hat{b}_i = 0.01\pm 0.05\}$, we may infer that the contamination from this particular template is additive, as $a$ and $b$ differ significantly from each other, and $b$ is consistent with 0 within the uncertainties. Fig.~\ref{fig: 25syst_cornerplot} shows a corner plot of a representative sample of estimated parameters constructed from the fitting in all 119 mocks for 6 of the 25 total systematics. 
%\rachel{This is figure 3, but before we were talking about figure 4.  Figures should be rearranged so they appear in the order they are discussed in the text.}\federico{Done} 
The blue (additive systematic) 
and red (multiplicative systematic) vertical and horizontal lines show the true parameter value. We see that on average, the combined model can recover the true parameters within the $1\sigma$ contours. Therefore, we can use the parameter estimates directly to learn about the forms of different systematics for which we have templates. Since this is done over multiple realizations, we use the parameter estimates for each mock to do this. However, in real data, we could use a bootstrap approach to generate similar cornerplots and study the contamination model from the data empirically. Using corner plots based on the MCMC parameters of a single realization is not equivalent to those produced from multiple realizations, as the full parameter variability is not captured by the MCMC error estimates.

\begin{figure*}
  \center
  \includegraphics[width=0.8\textwidth]{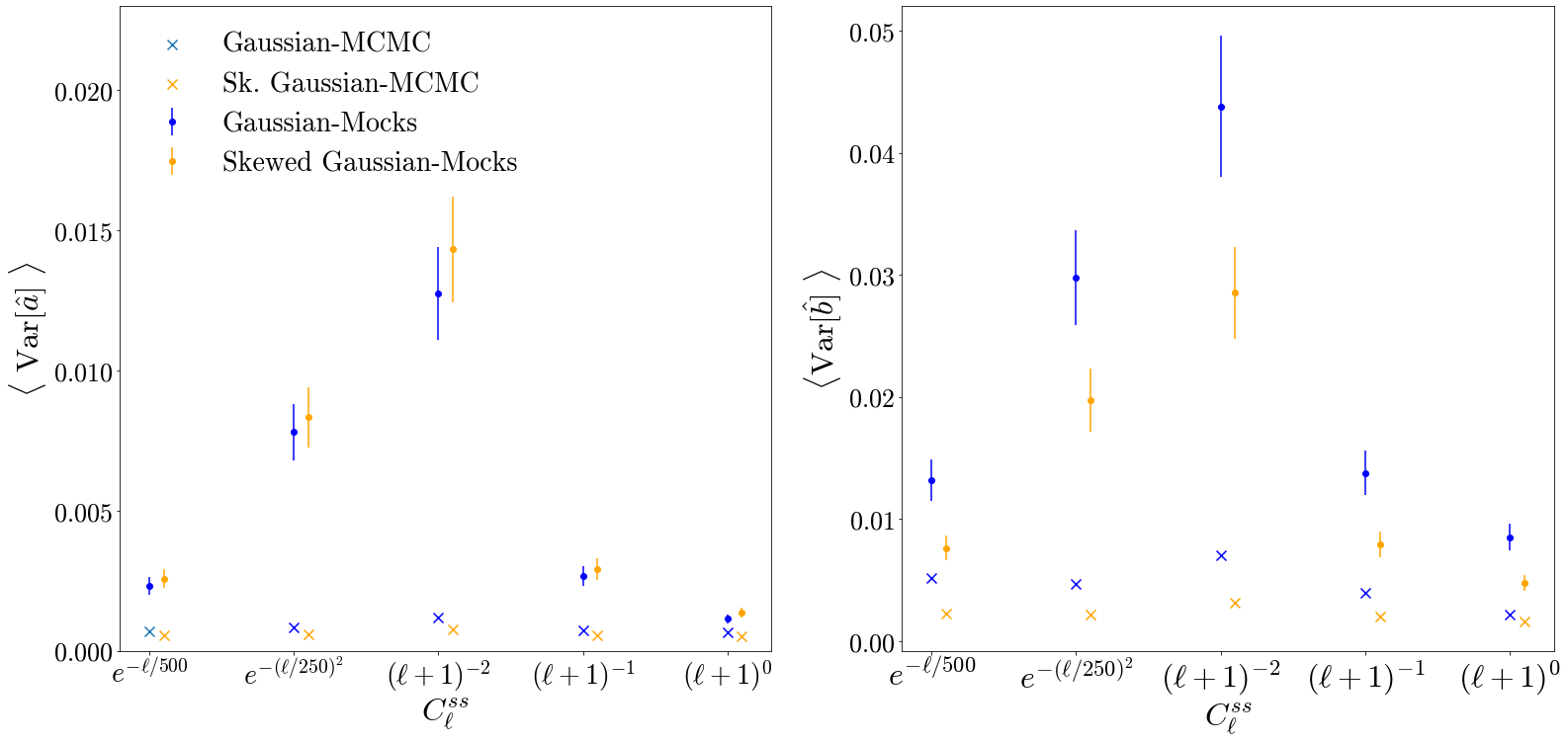}
  \includegraphics[width=0.6\textwidth]{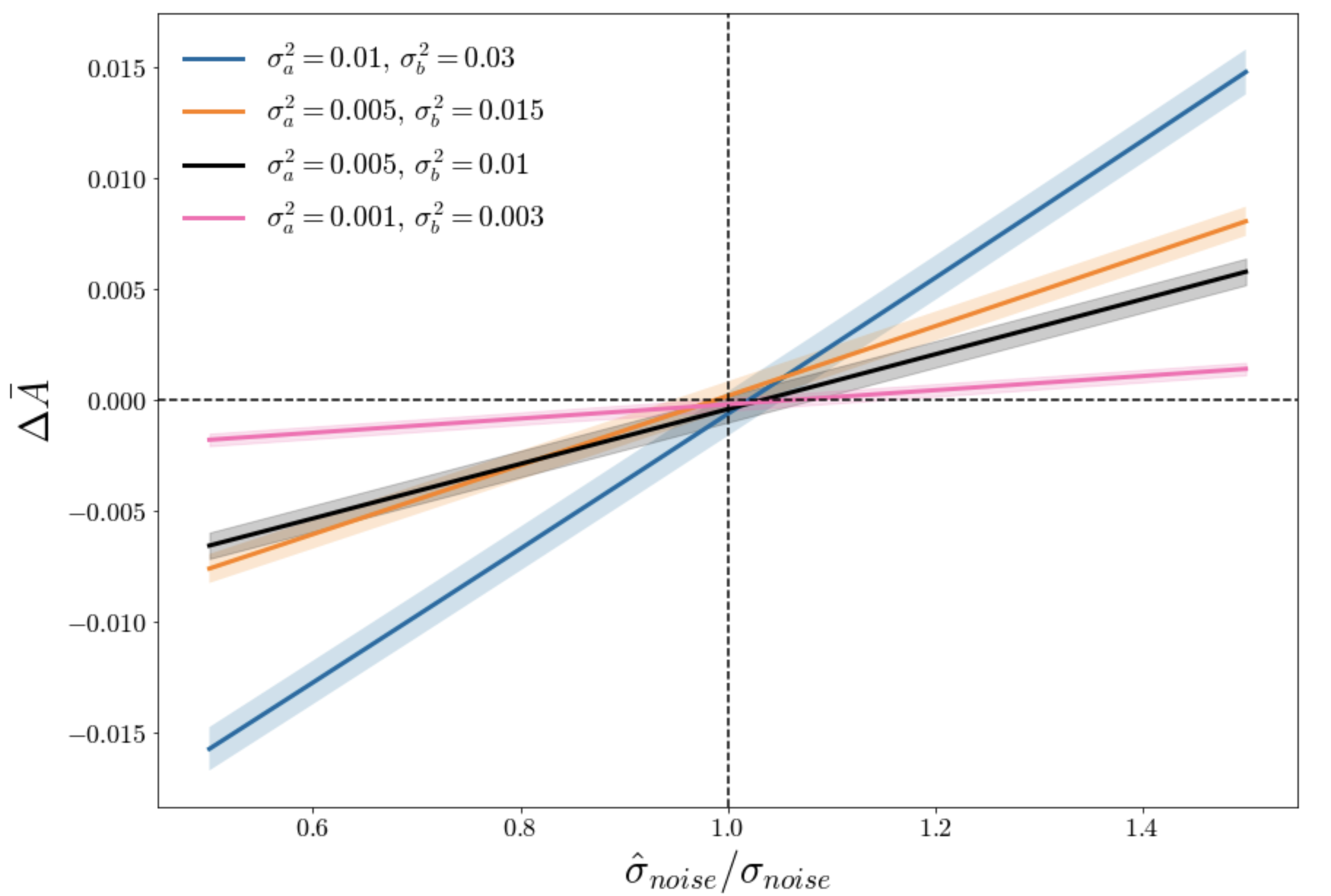}

  \caption{Top: Average parameter estimation variance for each systematics class used to generate contaminating templates. The variance for each estimated parameter is calculated over all 119 realizations of the KiDS-HOD mocks for each class of systematic. Since we use 5 systematics per class, the points show the average noise from the 5 parameters corresponding to each systematic contaminant from the mocks. The crosses show the parameter variance taken from the MCMC posteriors for comparison. The different colors show the likelihood form used, either a Gaussian (blue) or a Skewed Gaussian (orange). The errorbars are the errors on the mean. The left panel shows the variance on the $a$ parameters (which modify the mean of the observed $\delta_g$ distribution), while the right panel shows the variance for the $b$ parameters (which modify its variance). As we can see, there is little difference between the estimated noise using either likelihood on the $a$ parameters. However, the Skewed Gaussian has lower parameter noise for the $b$ parameters. We attribute this to the Skewed Gaussian's better recovery of the distribution's variance when we do not remove outliers. In addition, we see that the variance recovered from the MCMC posteriors vastly underestimates the parameter noise. Bottom: The effect of miscalculating the parameter noise on the two-point function correction. Possible parameters are drawn from a Gaussian centered at the true parameter value and with variance represented by the different colored lines. The correction to the observed correlation function is done using the estimated noise. The x-axis shows the ratio of the estimated noise to the true noise for both the $a$ and $b$ parameters, %\rachel{for both $a$ and $b$, i.e., we are misestimating the noise in both by the same factor?}\federico{Yes both by the same factor, I added a clarifying remark.}, 
  while the y-axis shows the bias in the corrected two-point function.  We see that under- (or over-) estimating the parameter noise causes the two point function to be over- (or under-) corrected. However, the effect of the bias depends on the size of the noise itself. Having high parameter noise means the analysis is more sensitive to accurately estimating the noise, while the opposite is true at low parameter noise. %\rachel{The x axis label on the top two plots uses $C_\ell^{ss}$, but the text uses $C_\ell^{tt}$ -- need to make figure consistent with text.} \federico{Addressed} 
  %\rachel{Can you remind me why we are plotting $\Delta A/\sigma_A$ rather than just $\Delta A$?  I think if it's a $5\sigma$ bias, we also want to know whether it's a 1\% or 5\% or whatever correction.  It seems to me that we can show both by plotting $\Delta A$ and using a shaded region to show $\sigma_A$.}\federico{Addressed. I think you are right this plotting $\Delta A$ makes more sense, I don't remember why we had chosen to do $\Delta A/\sigma_A$ when we talked about it.} 
  }
  \label{fig: debias_effect}
\end{figure*}

\begin {table}
\begin{adjustbox}{width=0.8\columnwidth,center}
\begin{tabular}{ ||c|c|c||} 

\hline
Contamination Model & $\Delta \bar{A} \; (\times 10^{-2}) $ & $\sigma_{\bar{A}} \; (\times 10^{-2})$ \\
\hline
Uncorrected &  $4.0 $ & $0.0135$  \\
\hline
%\texttt{SkewGauss\tu True} &  $-0.060$ & $0.284$  \\
%\texttt{SkewGauss\tu Combined} & $-0.064 $ & $0.484$  \\
True &  -0.0123 & 0.0631  \\
Combined &  -0.0166 & 0.0861  \\
Multiplicative &  0.1123 & 0.0728  \\
Additive &  0.7214 & 0.0374 \\

\hline
\end{tabular}
\end{adjustbox}
\caption {Comparison of the estimated bias on the two-point function ($w_{\text{corr}}(\theta) = (1 + \Delta A)w_{\text{true}}(\theta)$) %\rachel{need to check/fix equation (see my comments on section 4 about sign, also I think you might be missing a bar on top of $A$ here and throughout the caption}\federico{Addressed. I don't think a bar is missing, since the fit is done for every mock, the equation here shows the fit for each A, which is then averaged to obtain $\bar{A}$. In the caption I refer to A below by specifying I am talking about the mean of $A$, which is why I didn't include the caption.} 
for different assumed contamination models used when correcting the signal with 25 different systematics, including a mix of additive and multiplicative systematics. The first and second columns show the mean and error on the mean of $A$ over all mocks.  %\rachel{I'm not sure this is the most useful thing to show.  In fact, the most useful thing to show is $\Delta \hat{A}$ divided by the standard error on the mean value of that quantity -- in other words, how significant is the estimated bias?  From figure 4, this should be an extremely large number for the uncorrected signal, or for the two with incorrect model assumptions.  Perhaps we should talk this table through the next time we meet?  }
%\federico{Addressed based on discussion during meeting. Let me know if you think the changes help. There are still references in the text to the old notation, using the standard error. If this looks good I can erase them.} \rachel{Yes, please erase them.}\federico{Addressed} 
The uncorrected measurement is shown as a baseline for what our method needs to accomplish given the extremely statistically significant bias. %\rachel{How can we see this from the table, with the two entries 1.484 and 0.005?  As a rule, if you want somebody to conclude something from a table, it should be a straightforward number that is easy to see in the table.  I encourage you to reformulate the table such that this conclusion is obvious.  the rest of the caption likely needs some editing, but I'd like to wait until the table is fixed before doing so.}\federico{Addressed: the 16$\sigma$ is a typo from a previous incorrect version of the table.} 
Using the true contamination model (the correct choices of which systematics are additive or multiplicative) or the combined model both reduce the bias considerably (to below $0.1 \sigma$), while incorrectly applying the additive and multiplicative models to all templates results in a biased correction. The bias-variance trade-off can be observed by comparing both columns. The increased complexity of the combined model results in the highest variance, but in a considerable decrease on the bias when comparing with the additive and multiplicative models. The fitting results shown here use a Gaussian likelihood, but we confirm the differences with the skewed Gaussian form are small and within the errors from each other.}
\label{tab: bias}
\end {table}

We also note the tighter constraints on the $a$ parameters compared to the $b$ parameters. This discrepancy is due to the difference in constraining the shifts in the mean of the overdensity distribution (determined by the $a$ parameters) versus shifts in the variance (determined by the $b$ parameters). This can be understood statistically, as estimating the variance of the distribution intrinsically depends on estimating the mean. Any noise in estimates of the mean will propagate to estimates of the variance.

As shown here, the introduction of the combined model allows for joint modeling and correction of a set of clustering systematics that have different functional  forms.  Similarly, by construction it will allow for modeling of systematics that are neither purely additive nor purely multiplicative in overdensity.

\begin{figure*}
  \center

  \includegraphics[width=0.85\textwidth]{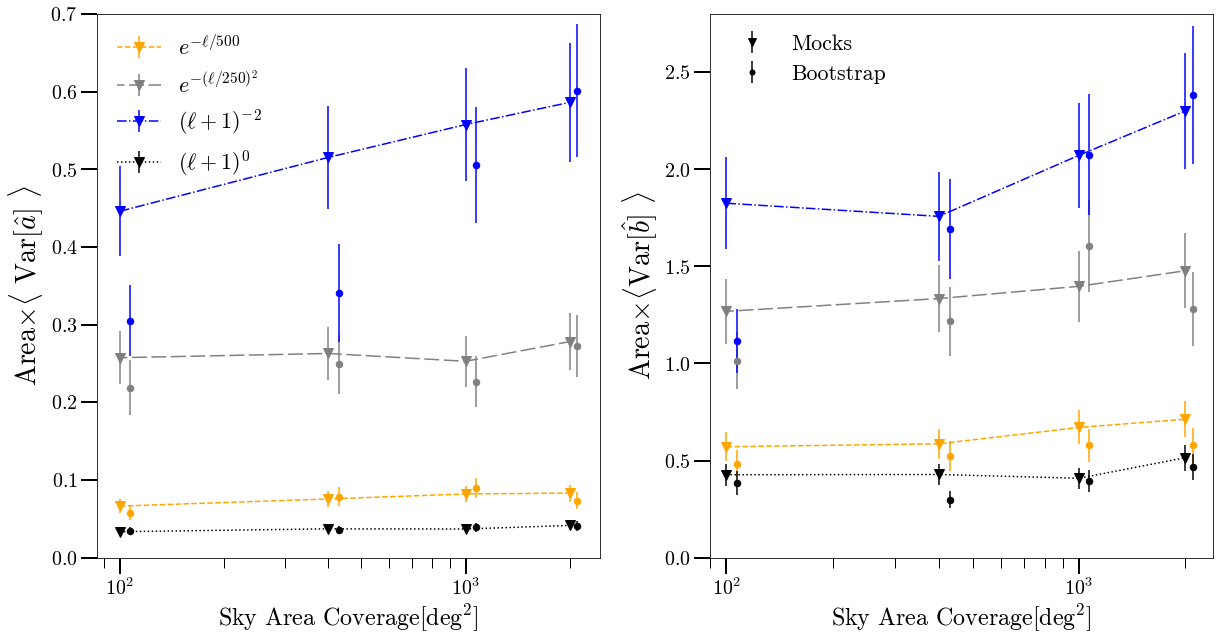}
  
  \caption{Comparison of the results of estimating systematics parameter noise using  multiple realizations of mock catalogs versus a block bootstrap approach on a single mock catalog realization. The figure shows the variance in the estimated parameters (scaled by the area) as a function of the sky area coverage. %We expect the parameter noise to decrease as our survey size increases. 
  The different line colors represent different systematic classes. The triangle points show the variance from using 100 realizations of mock catalogs, while the circular points show the variance estimated using the block bootstrap on a single realization. The errorbars show the error on the mean, and the lines (varying styles) track the results from mocks. We see that the bootstrap approach can effectively estimate the parameter noise for higher sky area coverage and low noise parameters, while it fails for small surveys with high parameter noise. %\rachel{Can you make the points larger so it's easier to see what is a triangle vs. circle?}\federico{Addressed}
  }
  \label{fig: noise}
\end{figure*}

\subsection{Cosmological Impact}\label{results: cosmo_impact}

To quantify the potential cosmological impact of how our method performs with multiple systematics, we fit the uncorrected and corrected clustering signals (for different modeling choices) to a constant times the true signal, as expressed in equation~\eqref{eq: corr_bias}.
Following the prescription detailed in Sec.~\ref{data and software: clustering_fitting}, we calculate the estimated bias, $\Delta \bar{A}$, for different systematics contamination models in the multiple systematics case described above and report the results in Table~\ref{tab: bias}. Although not shown, the differences between the estimated biases from using a Gaussian or skewed Gaussian likelihood are negligible, leading to a preference for the simpler Gaussian model. The bias-variance trade-off can be evaluated by comparing both columns in this table. As shown, the contaminated signal exhibits a few-percent bias that is highly statistically significant. 
% Once you get beyond 10$\sigma$ I don't really see the value in quoting numbers like $296\sigma$.  We don't think we know the answer that precisely, and the point is mostly that it's enormously significant.
%\rachel{There are some issues with this table, which I believe is not helping you cleanly tell this story -- see caption for explanation of this remark.  Once that is addressed, some more editing might be needed in this paragraph.}\federico{Addressed in table} 
Applying a correction using the true and the combined contamination models both reduce the bias to well below $0.1 \sigma$, with the caveat that the error on the mean from the combined model is approximately 1.4 times larger than that of the true model (increased variance). As expected, using the true contamination model is the optimal approach when it comes to reducing the bias without much increase in the variance. However, the combined model still dramatically outperforms the other contamination models and produces a bias consistent with zero, indicating that it is a sufficient correction for a mixture of different types of systematic contaminants for cosmological analysis. 

On the other hand, assuming the additive and multiplicative models in the presence of a mixture of different types of systematics produces a significantly biased correction. The correction using the additive model performs substantially worse than with the multiplicative model by a factor of $\sim 4$. We can see why from Fig.~\ref{fig: 25syst_correction}, where the multiplicative model does better than the additive at small scales and much worse at large scales. However, due to the smaller errorbars at small scales, the fitting of $A$ using equation~\eqref{eq: chi_2} favors the scales at which the multiplicative model performs better, which explains the lower values of $\Delta \bar{A}$ for that model. The increased complexity of the combined model results in the highest variance, but in a considerable decrease on the bias when compared with the additive and multiplicative models. Thus, the bias-variance tradeoff in this case favors the combined model.

%\rachel{All of the above is true, but I think another important conclusion you haven't yet tried to explicitly make is about the sufficiency of the correction with the combined model for cosmological analysis.}\federico{Addressed at the end of previous paragraph, let me know if the conclusion is sufficient}

\subsection{Effect of Correlated Systematics}\label{results: corr_syst}

So far we have only showed the performance of this method for uncorrelated systematics. However, we know that in reality template maps are often correlated with each other. To test the performance of this method for the case of correlated templates, we introduce different levels of in-class correlation between systematics templates\footnote{The more challenging issue of systematics templates that correlate with the real cosmological density field is deferred to future work.}. That is, in the process of map production using \texttt{HEALPY}, we provide off-diagonal elements that describe the correlation of all maps with each other. We only add correlation between maps that are produced from the same power spectrum (what we mean by in-class correlation), where the off-diagonal elements are $\rho_{\text{corr}} C^{t_it_j}_{\ell}$ if the maps $t_i$ and $t_j$ belong to the same class, and $0$ otherwise. We consider this scenario for a range of values of $\rho_{\text{corr}}$.

In our specific scenario, we are dealing with 5 systematic classes with 5 maps each. For our test, every map belonging to the same class will have the same level of correlation, $\rho_{\text{corr}}$, with each other and $0$ with all other maps. We then follow the method described in Sec.~\ref{methods: syst_model} to deal with such cases. Fig.~\ref{fig: 25syst_bias} shows the average bias in the corrected correlation function amplitude as a function of the in-class correlation. We see that in the presence of correlated template maps, there is no loss in performance.
%% RM: I don't think we have to call out higher biases, given the errorbars -- have commented out some extraneous discussion.
%The higher biases for some values of $\rho_{\text{corr}}$ are due to the resulting higher variance in the corrected two-point function coming from using the combined model. We can see this in the figure directly by looking at the size of the errorbars. 
All points are consistent within $2\sigma$ with an unbiased correction of the two-point function. In addition, the results with the Gaussian and skewed Gaussian likelihoods are once again not statistically significant.

%% file: Discussion.tex
\section{Discussion} \label{discussion}

\subsection{Comparison with other methods}
 Our methodology has some features in common with previous work described in Sec.~\ref{background:2}, but also some distinct differences. %\rachel{Refer back to section 2.3 for specifics.}\federico{Addressed} 
 Unlike the methods of Mode Projection \citep{Press_1992} and Template Subtraction \citep{Ross_2011},   %\rachel{name them - it's good for a comparison section to be precise}\federico{Addressed}, 
 we explicitly account for multiplicative terms in the one-point function describing the observed galaxy overdensity in equation~\eqref{eq: gal_overdensity}. %\rachel{I don't really understand the next sentence. 
 %The third term is the additive one, and that typically isn't neglected.}\federico{Yes this should be referring to equation 11, not equation 12} \rachel{Your comment says it should refer to equation 11 but it is referring to equation 2 - please check/correct.}\federico{Addressed} 
It is common practice to neglect the third term in this equation and consider only the first order term. However, we show in this work how neglecting the presence of multiplicative bias can lead to non-negligible biases in the corrected correlation functions if they are present. %Recall from equation~\eqref{eq: corr_bias} that the multiplicative terms modulate the correlation function and the linear contamination.%\rachel{This statement is only relevant in cases where multiplicative terms exist and are significant.  Perhaps you could give a reminder here of why that might be the case?}\federico{Addressed}

The method presented in this paper also considers distinct parameters contributing to the additive and multiplicative contamination to the galaxy overdensity, allowing for empirical determination of the systematics model, rather than fitting a specific model (e.g., additive or multiplicative). This not only allows for the possibility of multiple sources of contamination for a given template, but also allows us to directly determine the functional form of the systematic contamination.  The contamination parameters are also fitted for jointly, rather than independently for every systematic. This differs from iterative methods, which clean contaminants one by one using a form of linear regression. These methods rely on null hypothesis testing during the iterative process to avoid biased mitigation of contaminants. Other methods like \texttt{ElasticNet} incorporate the impact of multiplicative effects by using a two-step process to update the covariance for the assumed multiplicative parameters. In contrast, our method deals with this problem by taking into account all systematics simultaneously in the fitting process and self-consistently incorporating the multiplicative effects into the likelihood. This joint estimation enables the discrimination of additive and multiplicative templates directly.  %\rachel{Have you actually shown in the paper that this matters?  I know you tested it at some point but I didn't think the paper demonstrates that doing this one by one causes biases results.}\federico{Correct, it is not in the paper but I did test it. I think the correct interpretation is that the iterative methods have null testing conditions they need to meet to stop the iteration and prevent biased corrections. I have rephrased the sentence a bit, let me know if it's better }

In our current implementation of the method, we chose to mitigate the systematics directly at the two-point level, which differs from the iterative regression \citep{Elvin-Poole} and \texttt{ElasticNet} methods \citep{Huterer} that mitigate at the map level either through map cleaning or pixel weights. %\rachel{Refer to specific methods.  It's good for a comparison section to be precise.}\federico{Addressed} 
In principle, our method allows for mitigation at the map level through cleaning of the observed overdensity map using equation~\eqref{eq: mult_syst}.  % \rachel{You say this as if it is obvious, but to me it's not obvious, and below you seem to indicate it's not obvious - so this text needs some work.}\federico{Addressed: Added some additional clarification on this point}, 
%it is important to note the difference between corrections at the map level and at the two-point level. 
One common feature among corrections at these different levels is that all require a noise debiasing process  for the reasons explained in Sec.~\ref{methods: noise_debiasing}. Regarding the use of pixel weights, the last term in equation~\eqref{eq:ng_background} when used with combined model prevents us from being able to cleanly express the contamination as a weighted correction for each galaxy. We acknowledge that mitigating at the two-point level requires calculation of the auto-correlation function for all systematics of interest, which is not as simple as using pixel weights. However, it is still possible to perform a pixel-weighted correction, similar to that of iterative methods or \texttt{ElasticNet}, if the parameter fits agree with an additive or multiplicative model. In that case, the weights can be calculated for all templates $t_i$ with contamination parameter $a_{t_i}$ for any given pixel $j$ as:
\be \label{eq: pixel_weights}
w_j = \frac{1}{1 + \sum_{t_i}a_{t_i}\delta_{t_i,j}}
\ee

Finally, our method allows for fitting the $\delta_g$ distribution using a Skewed Gaussian likelihood, which more closely resembles the real distribution. Even though we do not see any significant difference in systematics mitigation between methods after debiasing the parameters, we still provide an alternative fit to a distribution commonly assumed to be Gaussian for systematics treatment\cite[see examples of non-Gaussinity in][]{Rezaie_2021, Rezaie_2023}. %\rachel{Is it worth noting that?  I really think it is misleading to give any significant emphasis to the results without debiasing, as that implies it's an optional part of the method.}\federico{Yes good point, I removed this.}  
We emphasize the non-Gaussianity of the $\delta_g$ distribution in the fits for the true distribution in Sec.~\ref{results: 0_syst}. There, the parameter that describes skewness was estimated to be $\gamma \approx 5.1 \pm 0.2$, %\rachel{need another significant figure on the 5 for this to be consistent, i.e. $5.x \pm 0.2$}\federico{Addressed}, 
which represents a highly non-Gaussian distribution. We note that the specific level of non-Gaussianity will heavily depend on the pixelazion scheme and galaxy density. Despite the high statistical significance with which the non-Gaussianity of the $\delta_g$ distribution is detected, the effectiveness of systematics mitigation for the Gaussian likelihood indicates that for our scenario, model mis-specification of a non-Gaussian distribution as a Gaussian is not a problem when parameter noise is known or can be estimated accurately.  %\rachel{Still not sure the logic is quite right -- if we can debias effectively, then why are we emphasizing the non-Gaussianity as our final point?  We need the takeaway to be very clear.  Should we add a final sentence that says `Despite the high statistical significance with which the non-Gaussianity of the $\delta_g$ distribution is detected, the effectiveness of systematics mitigation for the Gaussian likelihood indicates that for our scenario, model mis-specification of a non-Gaussian distribution as a Gaussian is not a problem.'  Or was there a scenario where the non-Gaussianity actually mattered?}\federico{I like that last sentence you propose, I added it. The non-Gaussianity only seems to make a difference for the cases of noise estimation (see Figure 8). The parameter noise for the "b's" was lower for skewed Gaussian, meaning if you under estimate the parameter noise with this Gaussian you are prone to higher biases than with the skewed Gaussian. This difference is mentioned in the text describing Fig 8, so I don't know if its worth mentioning here. I mention this detail at the end of the sentence you suggested, which is added before you original comment. } %\rachel{This statement is only meaningful if it's statistically significant.  Need errorbars to make this assertion.}\federico{Addressed}

\subsection{Practical applications on real data}\label{discussion: practical_applications}

%\rachel{I suggest changing the subsection title.  You've already applied the method to LSS.  The distinction you really want to get at is applying it to real data.  Perhaps `Practical applications on real data'?}\federico{Addressed} mention this question of pixel scale vs.\ characteristic scale of systematics in the future work for applying the method

Although our demonstration of the method used simulated data, the method is easily transferable to real LSS data. One open question is the choice of the pixel scale in relation to the characteristic scale of the systematics. More specifically, what pixelization is sufficient to properly characterize the systematic maps and mitigate contaminants? This work has produced and measured synthetic contamination at the same pixel scale; however, systematic template maps may be produced at finer scales in real data. The impact of pixel scale choices on the corrected clustering signal for systematics maps with varying characteristic scales was not explored here, but should be considered when working with real data. 

We have also emphasized that it is important to quantify the noise in the estimated systematics parameters in order to accurately mitigate the contamination. This work used multiple realizations of mock catalogs for this purpose. For surveys that have mock catalogs with realistic clustering, such an approach would transfer over directly.  However, not all surveys will have suitable mock catalogs. To apply the method without many realizations of mock catalogs, we propose the use of data-driven methods to estimate the noise in the systematics parameters. %Even though using mocks might be more accurate, they can be computationally expensive to create and real systematics maps may not be Gaussian, which may require more careful treatment in order to accurately simulate the real systematics used in fitting. 

Before applying data-driven methods to estimate the noise in the estimated systematics parameters, it is important to understand exactly what the noise depends on and how to evaluate its impact on systematics mitigation. The noise can depend on many factors, including the sky area coverage, the $\delta_t$ distributions and power spectrum, and the form of the likelihood used to analyze the $\dob_g$ distribution (including the number of parameters being fitted). For example, the top panel of Fig.~\ref{fig: debias_effect} illustrates how the noise level in the inferred systematics parameters can vary based on the form of the likelihood (blue and orange points represent results with a Gaussian and skewed Gaussian likelihood, respectively) and on the spatial power spectra used to generate the systematic maps, normalized to a fixed variance in $\delta_t$. As we can see, the noise can differ greatly depending on the power spectrum used, with some forms increasing the estimated noise variance by  almost an order of magnitude. The choice of a Gaussian or skewed Gaussian likelihood does not significantly affect the noise in the $a$ parameters, which control shifts in the mean of the distribution.  However, the choice has a substantial impact on the noise in the estimated $b$ parameters, which modulate its variance.
%
%This result hints at the fact that both distributions don't vary too much when estimating parameters that affect the mean, but the Skewed Gaussian is more precise about estimating the variance modulating parameters. This makes sense given that we have showed that the Skewed Gaussian does a better job at picking up the skewedness of the true underlying distribution, which affects the estimates of the variance. 
Therefore, even though the choice of the likelihood (Gaussian or skewed Gaussian) does not significantly affect the systematics mitigation outcome, it is important to note that the noise estimates differ. This could be relevant in the scenario where the parameter variance cannot be accurately measured and corrected for, in which case it is helpful to adopt a model with smaller variance since the bias correction would be less sensitive to it. However, we found no correlation between the noise variance in the estimated systematics parameter and the true value of the parameter. %Meaning that the noise is independent of the true value of the contamination parameter.

% Note from Rachel: I do not find the paragraph below compelling, and commented it out.  I don't think we need to make an argument for the need to debias properly.  It's part of the method and we clearly have to do it right.
%Also note that in our specific validation tests, we used a very small patch of sky (100 deg$^2$), which is significantly smaller than the area of current and future surveys such as DES HSC, or LSST. So given the small patch size, debiasing was especially important for this set of mocks, as portrayed by the bottom panel of Figure \ref{fig: debias_effect}. The different colored solid lines show different levels of parameter noise for the $a$ and $b$ parameters assuming all systematics have the same level of parameter noise. We see the effect of misestimating the parameter noise highly depends on the true noise itself and how badly its under (or over) estimated. If the parameter noise is low, the debiasing step becomes less important so misestimating the noise has little effect on mitigating contamination on the two point function. However, if the noise level is high misestimating the noise can lead to highly biased corrections. We therefore highlight the importance of properly understanding the parameter noise when debiasing. 
%\federico{Changes start here}

In the absence of mock catalogs, we propose the use of the block bootstrap method to estimate the parameter noise. Block bootstrap means that the pixelated maps are divided into approximately equal area patches, where each bootstrap sample represents a version of the maps made from a random selection (with replacement) of the patches. This allows us to create multiple realizations from a single map and estimate the parameter noise one would get from having multiple independent mocks. We wish to test this data-driven approach on higher and more realistic area coverage than our current KiDS mocks. %\rachel{Does that mean all the numbers above for the configuration were actually for a larger area?  If so, I think you should say this bit about making larger area maps {\em first} and only then describe the configuration in terms of numbers of blocks etc.}\federico{Addressed. It's for 4 area configurations, one is the same size and 3 are larger. I moved the explanation of how the larger maps are made before talking about the configuration.} 
To do so we use the same mechanism to generate systematic template maps from the set of 5 systematic classes described in the multiple systematic case in Sec.~\ref{results: mult_syst}. In order to obtain different area coverage $\delta_g$ maps, we used version 2.3.0 of \texttt{CCL}\footnote{https://github.com/LSSTDESC/CCL} \citep{Chisari_2019}  %\rachel{need citation and version number}\federico{Addressed. Wasn't sure if version number should go in paranthesis with citation, footnote, or just in text.} 
to generate galaxy-galaxy power spectra using a vanilla $\Lambda$CDM cosmological model ($\Omega_m = 0.3,  \Omega_b = 0.05, \; \Omega_{\Lambda} = 0.7,\; \sigma_8 = 0.8, \; n_s = 0.96$) with galaxy bias equal to $1$ ($\delta_m = \delta_g $) to produce Gaussian galaxy overdensity maps. This cosmological model is similar (while not identical) to that used in KiDS mocks, with the exception that our galaxy bias is fixed to 1. We use a Gaussian $n(z)$ with mean redshift $\mu_z = 0.3$ and $\sigma_z = 0.05$ to resemble the redshift range used for the previous validation with the KiDS mocks. We choose 4 size configurations: 100 deg$^2$, 400 deg$^2$, 1000 deg$^2$, and 2000 deg$^2$, to show how this approach works at various sky area coverages.  %\rachel{What area was used? You've said it was larger, but not how large?}\federico{Addressed above. 3 are larger and 1 is the same as the KiDS-HOD mocks (100 deg$^2$)} 
Although not identical, the produced mocks provide a realistic test of the block bootstrap method for estimating parameter noise. We wish to compare the parameter noise estimation using this bootstrap approach with estimates using multiple realizations. We transformed the Gaussian maps into log-normal maps to more accurately portray a realistic underlying distribution. This is done by taking some Gaussian map $G$, shifting it using its variance: $S =  G - 0.5*\text{Var}[G]$, and transforming the shifted map to a log-normally distributed one: $L =\exp(S) -1$. For the tests described here the shift parameter is $0.5*\text{Var}[G] = 0.025$.

In this specific example we use 100 bootstrap-resampled datasets (meaning 100 distinct versions of a map from one realization), with 10 block bootstrap patches. %\rachel{10 deg$^2$?  If not, 10 what?}\federico{Yes this was not clear, I made some edits hopefully it helps. Bootstrap size of 10 means the maps are divided into 10 equal area patches}, meaning each map is divided into 10 equal area patches from which each bootstrap sample is constructed. 
We confirm that this number of realizations leads to 5\%-level convergence of the parameter noise. The number of resampled datasets and patches %bootstrap sample and block size 
should be chosen based on the number of parameters estimated and scales at which correlations between pixels is small. In this particular test we estimate 51 parameters and correlations become negligible at $\sim 1.5 $ degrees, which motivates our chosen configuration. %is why we choose a sample size of 100 and block size of 10.\rachel{how can I relate this 64 to the 10 indicated above?  64 what?}\federico{This was a typo the 2 numbers should be 10, I had changed it to 64 because I had tried doing the bootstraping with block size of 64 but it was not as good as with 10, so I decided to keep the results from 10 and forgot to change it back here}. 

In Fig.~\ref{fig: noise} we compare the accuracy of parameter noise estimation by comparing a bootstrap approach on a single mock catalog realization against our previous approach using multiple realizations of mock catalogs as a function of sky area coverage. As we can see from the figure, the noise is approximately proportional to the inverse of the survey area. %Therefore, we can expect the debiasing step in our method to have a lesser importance for surveys with larger area.  %% RM: don't say this, as we don't want to imply that we've identified a regime where it's unnecessary; also, larger area surveys have smaller statistical uncertainties, so all aspects of systematic mitigation are more important there
The figure also shows that using the block bootstrap on a single dataset can be a good alternative to using multiple realizations of mock catalogs. %\rachel{Did you check whether results had converged with 100 mock realizations? 
 %Once you use CCL, there is no reason not to check whether that's enough or whether more realizations are needed.  I think until you've done that, we can't really use the mocks as the idealized reference point.} \rachel{ Regarding the bootstrap, simply using 100 regions regardless of survey area is certainly not how people would do the analysis in reality, and it could be causing your results to look artificially bad for small survey areas.  Once your regions are small enough, the bootstrap assumption fails (since the galaxy clustering signal makes the correlations between the regions more important).  Most people would therefore use a fixed block size, or at least impose a minimum block size, rather than always using the same number of blocks for all areas.  I suggest modifying your approach to be more like what would be done in real data, and seeing whether the results change. 
% Otherwise people will not find the test compelling.  (I did not edit text below this in the paragraph because it might possibly change.)}\federico{I should have been more clear in the text. The block size is not 100 (the area is not divided into 100 regions), its much smaller ( 10 regions). 100 is the number of times this process of re sampling the regions is done, to obtain 100 different "mock" catalogs. Hopefully that makes it clearer. I checked convergence and 100 mocks might not be enough. Addressed mock convergence above} 
 The bootstrap estimates are quite good for higher area mocks, and for small area mocks with lower levels of parameter noise.  However, the approach fails for small area surveys ($\sim$100~deg$^2$) with high parameter noise associated with systematics maps that have a lot of small-scale power. %This highlights that the quality of the estimate can be sensitive on both the size of the noise and the survey coverage. 
 
 In addition, the bootstrap estimates can depend on of the number of patches chosen. Creating patches that are too small might fail to capture large spatial correlations between template maps. Conversely, choosing very few large patches can result in challenges in inferring a significant number of systematics parameters and their variances. Our results show, however, that it is possible to achieve a balance that reliably estimates the systematics parameter noise for many realistic survey scenarios.

Note that we cannot simply take the parameter noise variance directly from a single realization by looking at the posterior generated from the MCMC samples. Because of our simple adopted likelihood model, which neglects correlations in overdensity between pixels, the variance of the posterior coming from the MCMC samples does not reflect sources of uncertainty such as cosmic variance. Using the MCMC sample posterior noise will produce underestimates of the parameter noise, as seen in Fig.~\ref{fig: debias_effect}, and therefore lead to overcorrection of the two-point function. %So although tempting, we do not recommend the use of the MCMC posterior from a single run as a way to estimate the parameter noise. 

%% file: Conclusion.tex
\section{Conclusions} \label{conclusions}
In this paper, we presented a new method for characterizing and mitigating systematic contaminants in in galaxy overdensity fields. We used a generalized and flexible contamination model that allows for multiple systematics to contaminate the overdensity field in additive, multiplicative, or `combined' forms. This modelling choice allows us to quantify contamination without assuming that all systematics follow one particular form, and in addition helps us empirically determine the functional dependence of systematics on the galaxy overdensity. Using the generalized model can reduce model biases, producing unbiased estimates of the two-point correlation function in the cases where different systematics take different forms captured by our model. 
 The ability to determine the functional form of the contamination may provide insight into their origin, so they can be reduced or entirely avoided in future. %This work emphasizes the ability to empirically learn about the functional form of contaminants, as current and future surveys may present systematics of unknown contamination form. 

 As part of our approach, we analyze the one-point function of galaxy overdensities within pixels to determine the systematics contamination model, then use that to correct the two-point functions.  %\rachel{I would not say the next sentence as-is.  You need to provide a concise statement of what likelihood you recommend.  If you recommend the Gaussian likelihood because it performs just fine despite being less realistic, you should say that. If the skewed Gaussian is necessary for accurate systematics mitigation, you should say that.  But just the fact that you've made a skewed Gaussian likelihood is not information that is helpful to provide in the conclusion on its own without further context on its value.}\federico{Addressed} 
 In addition, we introduced a skewed Gaussian likelihood function to represent the galaxy overdensity distribution in an attempt to more closely model the true underlying conditional probability function. However, there were no significant differences in the quality of systematics mitigation between the skewed Gaussian and Gaussian likelihood. Therefore, we recommend using the simpler Gaussian likelihood. 

The method was tested and validated using KiDS-like mock galaxy catalogs with realistic clustering, and mock systematic templates maps. Template maps with different systematic `families' were produced to mimic the level of complexity we can expect to find in real data. Key outcomes of our tests were as follows:

\begin{enumerate}
    %\item \rachel{I am not sure this bullet point is needed.  Your focus at this point should be how well does your method mitigate systematics. This bullet point does not get you there.}\federico{Adressed, commented out bullet point} No systematics: We tested the assumption of Gaussianity of the $\delta_g$ field under no contamination and found that the skewed Gaussian distribution provides a better fit to the data than the more commonly used Gaussian distribution, motivating our choice of likelihood parametrization.

    \item Contaminating the true $\delta_g$ with one template map in two different ways (multiplicatively and additively), we found that applying the wrong model when mitigating systematics can result in significant residual biases after correcting the two-point function, while our proposed, more general model produced unbiased corrections and has comparable performance to the  case where the true contamination model is known. We also showed that simple likelihood ratio tests can successfully identify the correct contamination model. Both the Gaussian and  skewed Gaussian likelihood model for the overdensity field result in robust systematics mitigation, leading us to prefer the simpler Gaussian model for correction purposes even though it does not describe the overdensity field particularly well.

    \item Multiple systematics: We contaminated the overdensity field with a non-trivial number of contaminants (25) with different power spectra, of which some were additive and some multiplicative systematics. We showed that the combined model once again produced unbiased estimates of the clustering signal, while assuming a single contamination model for all systematics produced biased corrections in this mixed case. In addition, we found that we can accurately determine the functional form of the systematic contamination for the individual systematics.
\end{enumerate}

%We also discussed the differences between the proposed method and already existing mitigation techniques, highlighting the differences in the contamination model, ability to empirically study the functional form of contamination, and our choice to mitigate directly at the two-point level rather than at the map level.
%\rachel{I'm not sure how helpful the previous sentence was, because you haven't given any outcomes; people would have to go back and read the relevant section to get anything from it.  I would suggest either omitting the previous sentence, or providing a few key takeaways from the comparison.} \federico{Addressed: omitted previous sentence}
Finally, we considered the practical applications of the methods presented here in real data. The main challenge is to accurately estimate the systematics model parameter noise needed to obtain unbiased corrected  clustering signals. Since this work used mock catalogs,  parameter noise could be estimated well from multiple mock catalog realizations. However, this will pose a challenge when the method is applied to real data, as mock catalogs are not always available and are computationally expensive. To overcome this challenge, we proposed a block bootstrap approach for estimating the parameter noise from a single realization of both the galaxy density field and systematic maps. We showed that the estimated parameter noise using the block bootstrap is comparable to the estimates using multiple mock catalog realizations in all cases except for small area ($\sim$100~deg$^2$) with significant small-scale power in the systematics template maps. This is therefore a promising approach in applying this method to real data in future work. 

Applications of this method on current imaging surveys would allow us to directly compare the results with commonly used mitigation techniques on real data. In addition, empirical determination of the functional form of real systematics could help us learn how systematics bias the observed galaxy overdensity field. Recent analyses like the DES Y3 clustering analysis \citep{Rodriguez_Monroy_2022}, where concerns emerged regarding large-scale structure systematics,  are an interesting and highly relevant context for direct application of the method developed in this work, as a stepping stone towards future surveys. %Additional tests on the bootstrap approach to estimate parameter noise is needed to properly understand it's accuracy in real data applications. 

%Additional understanding of  